\documentclass[fleqn,usenatbib]{mnras}

\usepackage[T1]{fontenc}

\DeclareRobustCommand{\VAN}[3]{#2}
\let\VANthebibliography\thebibliography
\def\thebibliography{\DeclareRobustCommand{\VAN}[3]{##3}\VANthebibliography}

\usepackage{graphicx}   % Including figure files
\usepackage{tabularx}
\usepackage{amsmath}    % Advanced maths commands
\usepackage{amssymb}    % Extra maths symbols

\usepackage{newtxtext,newtxmath}

\newcommand{\bs}{\boldsymbol}
\newcommand{\percent}{\,\mathrm{per\,cent}}
\usepackage{xcolor}
\definecolor{pink}{rgb}{0.96, 0.76, 0.76}
\definecolor{aqua}{rgb}{0.22, 0.96, 0.93}
\definecolor{darkred}{rgb}{0.76, 0.23, 0.13}

\title[The Gaia DR3 Mira variable period--age relation]{A kinematic calibration of the O-rich Mira variable period--age relation from Gaia}

\author[H. Zhang \& J.~L. Sanders]{Hanyuan Zhang
and
Jason L. Sanders\thanks{hz420@cam.ac.uk, jason.sanders@ucl.ac.uk}
\\
Department of Physics and Astronomy, University College London, London WC1E 6BT, UK\\
}

\date{Accepted XXX. Received YYY; in original form ZZZ}

\pubyear{2023}

\begin{document}
\label{firstpage}
\pagerange{\pageref{firstpage}--\pageref{lastpage}}
\maketitle

\begin{abstract}
Empirical and theoretical studies have demonstrated that the periods of Mira variable stars are related to their ages. This, together with their brightness in the infrared, makes them powerful probes of the formation and evolution of highly-extincted or distant parts of the Local Group. Here we utilise the Gaia DR3 catalogue of long-period variable candidates to calibrate the period--age relation of the Mira variables. Dynamical models are fitted to the O-rich Mira variable population across the extended solar neighbourhood and then the resulting solar neighbourhood period--kinematic relations are compared to external calibrations of the age--kinematic relations to derive a Mira variable period--age relation of $\tau\approx(6.9\pm0.3)\,\mathrm{Gyr}(1+\tanh((330\,\mathrm{d}-P)/(400\pm90)\mathrm{d})$. Our results compare well with previous calibrations using smaller datasets as well as the period--age properties of Local Group cluster members. This calibration opens the possibility of accurately characterising the star formation and the impact of different evolutionary processes throughout the Local Group.
\end{abstract}

%%%%%%%%%%%%%%%%%%%%%%%%%%%%%%%%%%%%%%%%%%%%%%%%%%

%%%%%%%%%%%%%%%%% BODY OF PAPER %%%%%%%%%%%%%%%%%%
\section{Introduction}

In the study of the formation and evolution of the Milky Way, one crucial ingredient is accurate stellar ages \citep{FreemanBlandHawthorn2002,BlandHawthornGerhard2016}. With this information, we can begin disentangling the series of events that have led to the observed Milky Way today, as well as directly measure the dynamical restructuring of the Galaxy. However, despite their clear advantages in analysing the Galaxy, stellar ages are awkward quantities due to their indirect measurement only via stellar models. Many stellar age indicators exist \citep{Soderblom2010} which often provide different levels of accuracy for different stellar types and different stellar populations. With the availability of Gaia astrometry \citep{Gaia1,Gaia2} and complementary large-scale spectroscopic surveys \citep[e.g.][]{deSilva2015,Majewski2015}, two methods applicable to large collections of stars are comparisons to isochrone models \citep[e.g.][which operates most successfully for subgiant stars that have recently turned off the main sequence]{Xiang2017,SandersDas2018,XiangRix2022}, and indirect mass measurements of giant stars through spectroscopic measurements of the products of dredge-up episodes calibrated via asteroseismology \citep[e.g.][]{Masseron2015,Martig2016}.

Mira variables are high-amplitude thermally-pulsing asymptotic giant branch (AGB) stars. Their study in the Large Magellanic Cloud \citep[e.g.][]{GlassLloydEvans1981,Wood1999,Groenewegen2004} demonstrated that they follow a tight period--luminosity relation (believed to be associated with fundamental mode pulsation) making them interesting tracers both for local Galactic and cosmological studies \citep{Catchpole2016,Grady2019,Grady2020,Huang2020}. The chemistry of Mira variables is either oxygen or carbon-dominated depending on the C/O ratio \citep{Hofner2018}, but O-rich Mira variables are significantly more common in the Milky Way and are found to follow tighter period--luminosity relations due potentially to less circumstellar dust \citep{Ita2011}. It has long been empirically known that groups of Mira variables binned by period show correlations between period and scaleheight/velocity dispersion \citep{Merrill1923,Feast1963}, which is typically interpreted as a correlation between the period and age of a Mira variable where the older stars have longer periods. This opens the possibility of using Mira variables as age indicators within the Galaxy and beyond \citep[e.g.][]{Grady2020}. A limited number of Mira variables in clusters also validate the period--age connection although confident assignment of membership has only been possible recently with Gaia data \citep{Grady2019,Marigo2022}. Although the period--age relation has been approximately calibrated empirically \citep{FeastWhitelock2000}, relatively few theoretical models reproducing the behaviour exist \citep{WyattCahn1983,Eggen1998,Trabucchi2022} and the lack of detailed reproduction of the period--luminosity relations of fundamental mode pulsation from theoretical models suggests the period--age relations still have some associated uncertainty and there is a need for accurate data-driven calibrations.

Encounters in the stellar discs of galaxies cause stellar populations to slowly kinematically heat giving rise to age--velocity dispersion relations
\citep{Wielen1977} such as those suggested for Mira variable stars \citep{Feast1963}. There are multiple suggested perturbers that give rise to disc heating including molecular clouds, spiral arms or merger events \citep{SS1,SS2,Barbanis1967,Velazquez,Hanninen2002,Aumer2016} that likely have differing relative contributions across the Galactic disc \citep{Mackereth2019}. In the solar neighbourhood, the stellar velocity dispersion is approximately a power law in age with exponent $\sim0.3$ for the radial dispersion and $\sim0.5$ for the vertical dispersion \citep{Holmberg2009,AumerBinney2009,Sharma2021}. A common picture \citep{BT08} for this behaviour is that the spiral arms are efficient in-plane heating sources giving rise to the increase in radial velocity dispersion and molecular clouds are efficient in converting this radial energy into vertical energy \citep{Aumer2016}. There is the further complication that the stellar populations could have been born hotter in the past, which could play a part in the observed correlations \citep{Bird2021}. Now with Gaia data, the age--velocity dispersion relations can be inspected across the Galactic disc \citep{SandersDas2018, Mackereth2019, Sharma2021, Antoja2021}.
For our purposes, the fact that correlations between age and kinematics exist is sufficient and we need not necessarily understand the underlying cause. In this way, kinematics can be used as an age proxy for groups of stars. Note that for this procedure to operate well, we are perhaps implicitly assuming that the kinematic--age relations are monotonic as evidenced in the solar neighbourhood \citep[e.g.][]{Holmberg2009}.

With the publication of large catalogues of variable stars from Gaia with associated proper motions \citep{Eyer2022}, there is now the possibility of thorough characterizations of the dynamical properties of different families of Mira variable stars \citep{Alvarez1997}. Kinematic characterization then opens up the possibility of mutual age calibration of different age tracers. By assuming kinematics are solely a function of age, we can anchor different age indicators to each other by requiring they all reproduce the same age--kinematic relations \citep[e.g][]{Angus2015,Angus2020}. In this way, we can characterise the Mira variable period--age relation. This simplifying assumption can be complicated by metallicity dependence, particularly if different tracers are biased toward different metallicity populations. The Mira variable stage occurs in stars of all metallicities although C-rich Mira variables are only formed through dredge-up in young, metal-poor stars \citep{Boyer2013}. This strategy of mutual age calibration via age--kinematic relations has been utilised successfully in the study of gyrochronology \citep{Angus2015} and chromospheric activity in late-type stars \citep{Wilson1970,West2015}, and promises a route to the mutual calibration of all stellar age indicators.

In this work, we utilise the astrometry of the latest Gaia DR3 long-period variable candidate catalogue to characterise the kinematic behaviour of O-rich Mira variables separated by period and combine this information with literature age--velocity dispersion relations in the solar neighbourhood to characterise the period--age relation for O-rich Mira variable stars. In Section~\ref{sec::data} we describe the dataset we use focusing on the cuts required to isolate both O-rich AGB stars and those high-amplitude long-period variables that are likely Mira variables. In Section~\ref{sec::model} we describe our modelling procedure and tests on mock data, before showing the results applied to data in Section~\ref{sec::application} and the resulting period--age relation in Section~\ref{sec::period_age_relation}. We critically discuss our approach and compare to other Mira variable period--age relations in Section~\ref{sec::discussion} before summarising our conclusions in Section~\ref{sec::conclusions}.

\section{The Gaia DR3 O-rich Mira variable sample}\label{sec::data}
\begin{figure*}
    \centering
    \includegraphics[width = \textwidth]{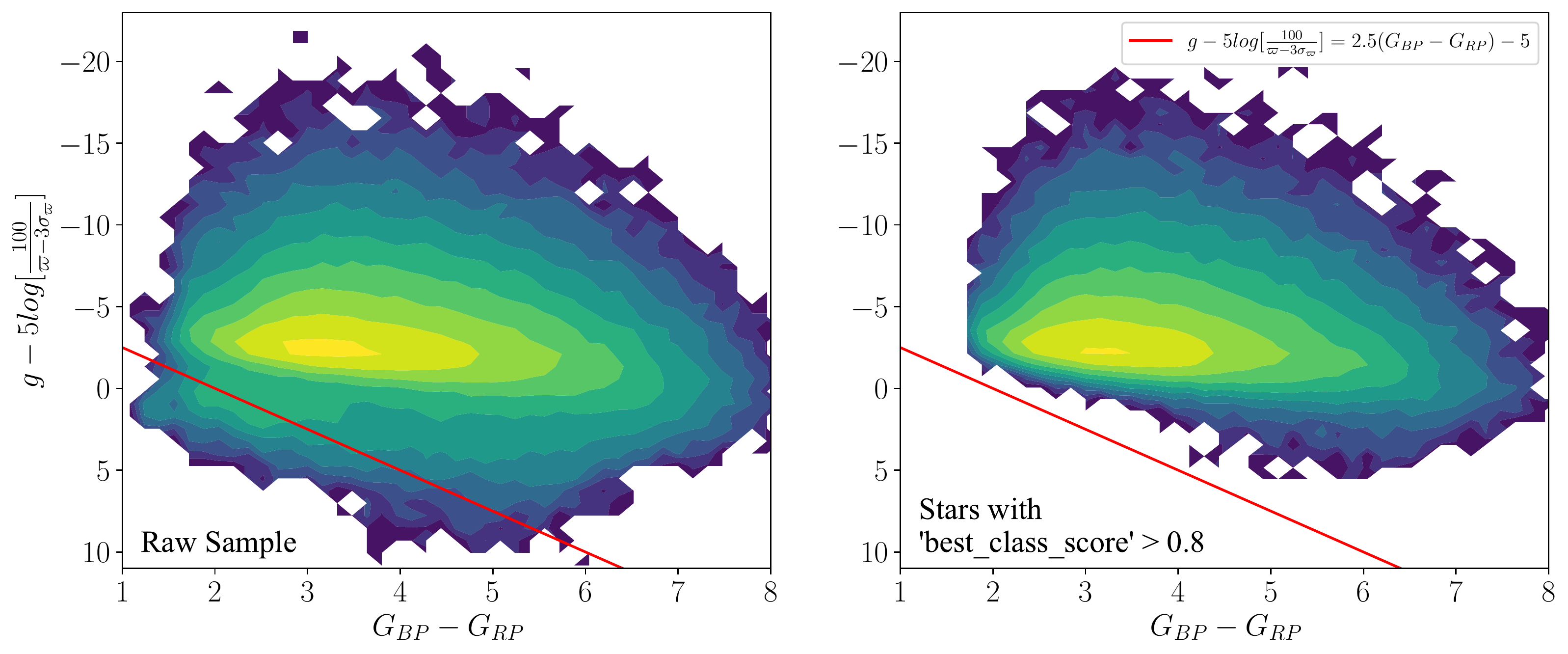}
    \caption{Colour--magnitude diagrams computed using a $3\sigma$-adjusted parallax, $\varpi-3\sigma_{\varpi}$. We define the region occupied by AGB stars as $G - 5\log_{10}(100/(\varpi-3\sigma_{\varpi})) < 2.5(G_\mathrm{BP}-G_\mathrm{RP})-5$: any star outside this is likely a YSO. The right panel shows those only those stars with $\texttt{best\_class\_score}>0.8$ which effectively removes any likely YSO contaminants.}
    \label{best_class_score}
\end{figure*}

\begin{figure}
    \centering
    \includegraphics[width = \columnwidth]{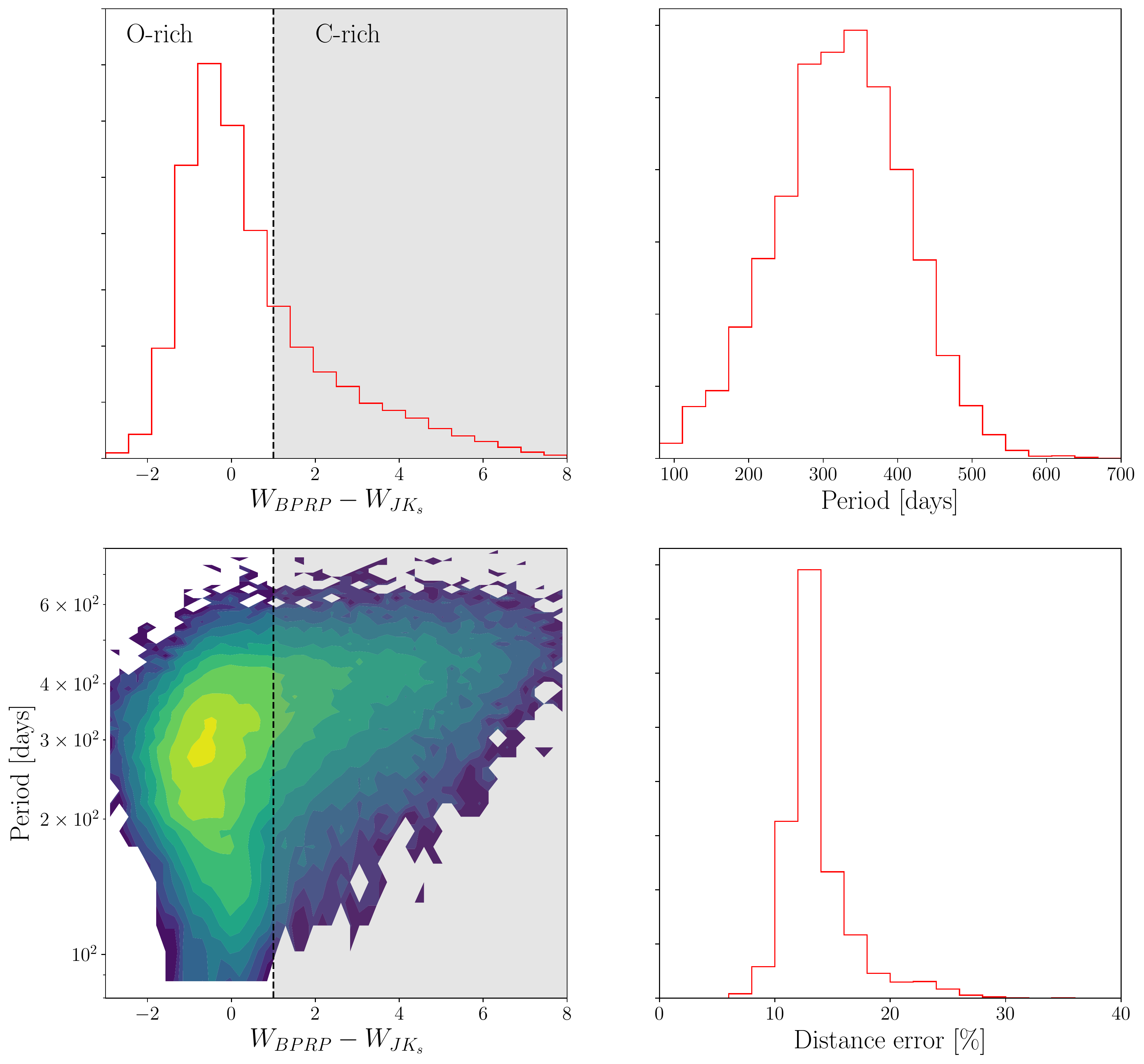}
    \caption{Properties of our O-rich Mira sample: the top left panel shows the distribution of the Wesenheit index difference from \protect\cite{Lebzelter2018} used to separate O-rich and C-rich Mira. The lower left panel shows the distribution of this quantity vs. period. The right two panels show the period and distance error for the O-rich Mira sample.}
    \label{orich-crich_property}
\end{figure}

\begin{figure*}
    \centering
    \includegraphics[width=.49\textwidth]{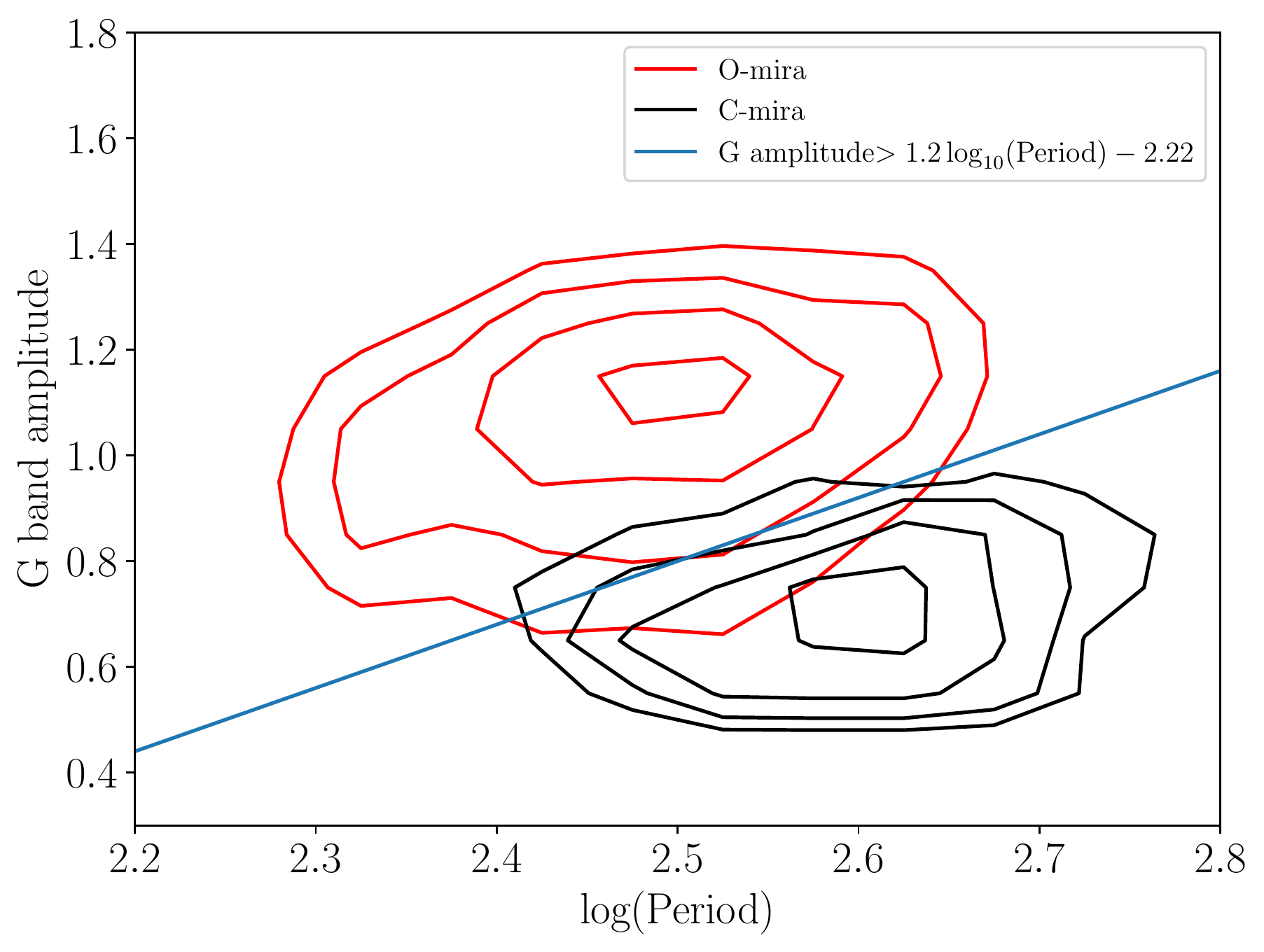}
    \includegraphics[width=.49\textwidth]{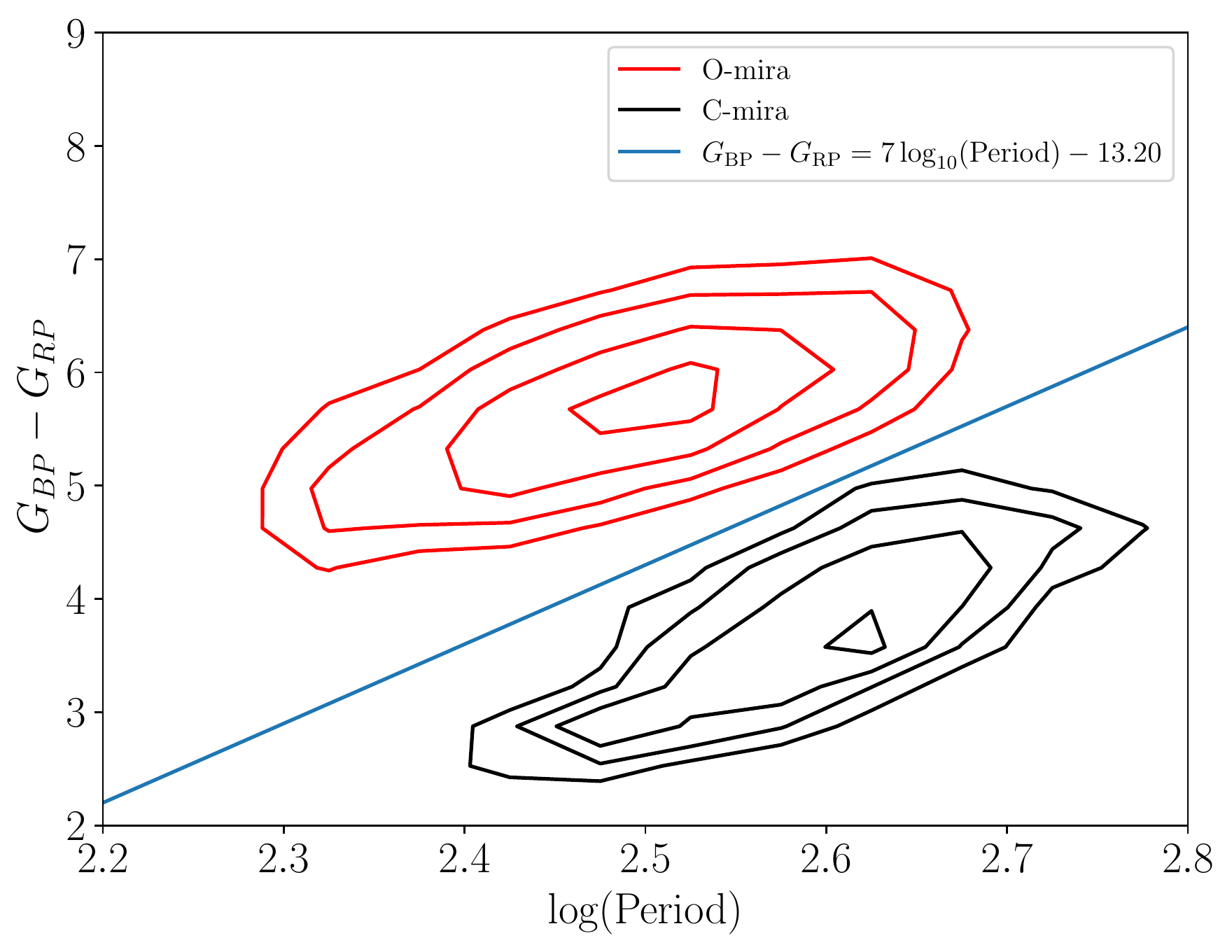}
    \caption{The contour plot of the C-rich (black) and O-rich (red) Mira variable population selected by their spectrum on period--amplitude plane and period--colour plane respectively. Candidates below the blue line were removed from the sample.}
    \label{selection_O_C_mira}
\end{figure*}
We begin by describing how we form our O-rich Mira variable sample. It is important to note that our analysis relies on characterising the velocity distributions at each Galactic location. In this way, considerations on the completeness of our sample are unimportant provided we do not perform any specific selections on the velocities of the stars. Our primary objective with the selection is to form a low-contamination subset.

We use the long period variable (LPV) candidate catalogue from Gaia DR3 \citep{Lebzelter2022}. This catalogue has been constructed in a two-stage process -- likely variable stars are identified by comparison to literature variable sources and reference non-variable Gaia sources, and then classified based on literature classifications and features including the Lomb-Scargle period, time summary statistics, colours and parallax \citep{Holl2018,Rimoldini2019,Rimoldini2022}. Stars classified as LPVs with $G$ $5$th$-95$th percentile greater than $0.1\,\mathrm{mag}$ and $G_\mathrm{BP}-G_\mathrm{RP}>0.5$ (along with other less important cuts for our purposes) were further considered by the specific object study (SOS). Candidate LPVs from the SOS have published generalised Lomb--Scargle periods (and Fourier amplitudes) in Gaia DR3 if the period is greater than $35\,\mathrm{day}$ and shorter than the $34$ month time series duration, the $G$-band signal-to-noise $>15$ and there is no correlation between the image determination parameters and the time series. Infrared photometric measurements were acquired from the 2MASS catalogue \citep{Skrutskie2006} using the precomputed cross-match provided on the Gaia archive. There are $1\,657\,987$ variable star observations in the Gaia DR3 LPV candidate SOS catalogue after the cross-match with 2MASS. We first remove stars without measured periods or without $J$ and $K_s$ photometric measurements which are needed for later selection pipelines. These requirements reduce the size of the sample to $387\,419$ objects. 

To isolate a sample of likely Mira variables, we employ cuts in period and magnitude. We retain stars with $80<\mathrm{Period}/\,\mathrm{day}<1000$ \citep{Matsunaga2009} and in amplitude we employ a similar cut to \cite{Grady2019}, which removes stars with $\texttt{amplitude} < 0.5\,\mathrm{mag}$ (compared to \citealt{Grady2019} cut at $0.43\,\mathrm{mag}$). Here \texttt{amplitude} is the $G$ semi-amplitude computed from a Fourier fit. Note that around the problematic period of $190$ day, the Fourier fit can significantly overestimate the amplitude of the LPVs leading to lower-amplitude semi-regular variable contaminants in a Mira variable selection. We remove stars with $170<\mathrm{Period (days)}<200$ and $\texttt{amplitude}>1.3$, and $350<\mathrm{Period (days)}<400$ and $\texttt{amplitude}>1.6$ to mitigate against this.

As highlighted by \cite{Mowlavi2018}, young stellar objects (YSOs) can be a contaminant in the LPV processing as they have similar colours, amplitudes and periods to LPVs. In the classification pipeline from \cite{Holl2018} and \cite {Rimoldini2019}, the probability of the object being of the reported class, $\texttt{best\_class\_score}$, seems an effective indicator of YSOs. In Fig.~\ref{best_class_score}, we show the colour--absolute magnitude diagram for our sample computed using a parallax adjusted by $3$ times the parallax uncertainty. This gives the brightest possible magnitude for each star within the parallax uncertainties so any star consistent with being near the main sequence using this measure is likely a YSO. Many of these objects also have $\texttt{best\_class\_score}<0.8$ so we choose to only consider stars with $\texttt{best\_class\_score}>0.8$. From this series of cuts, we end up with $75\,874$ Mira variable star candidates.

\subsection{O-rich/C-rich classification}
LPVs can be either oxygen-rich or carbon-rich depending on the metallicity and the strength of the dredge-ups which is controlled by the initial mass \citep{Hofner2018}. The O-rich stars follow a tighter period-luminosity relation \citep[due to increased circumstellar dust in the C-rich stars,][]{Ita2011} and are significantly more common in the Milky Way \citep[with C-rich stars contributing more in the outer disc,][]{Blanco1984,Ishihara2011}.
As shown by \cite{Lebzelter2022}, the Gaia DR3 BP/RP (XP) spectra can be used to effectively separate O-rich and C-rich AGB stars due to the differing set of band heads and features in their spectra arising primarily from the TiO and CN absorption features. Sanders \& Matsunaga (submitted) have provided an unsupervised classification approach for these spectra that effectively separates O-rich and C-rich LPV stars and performs better than the Gaia DR3 classifications for highly-extincted sources. We adopt their classifications where Gaia DR3 XP spectra are available. \cite{Lebzelter2018} showed that, within the LMC, O-rich and C-rich Mira variables can be separated in the plane of $W_\mathrm{BPRP}-W_{JK_s}$ vs. $K_s$. Here the two Wesenheit indices are $W_\mathrm{BPRP}=G_\mathrm{RP}-1.3(G_\mathrm{BP}-G_\mathrm{RP})$ and $W_{JK_s}=K_s-0.686(J-K_s)$. Although the boundary employed by \cite{Lebzelter2018} is slightly curved, we can employ a very similar cut to select O-rich Mira as $W_\mathrm{BPRP}-W_{JK_s}<1$. The left two panels of Fig.~\ref{orich-crich_property} show that this Wesenheit index difference against period for the selected Mira sample, whilst the right panels are the period and distance percentage error of the O-rich Mira after further selections. The performance and purpose of these two cuts are very alike, but we employed both cuts here to maximally remove C-rich Mira contamination.

Aided by the XP spectrum classifications, we have found that O-rich and C-rich sources are separated in the period--amplitude plane and period--colour plane as shown in Fig.~\ref{selection_O_C_mira}. Hence, we make a further two cuts to remove those C-rich Mira variables when an XP classification is not available: $\texttt{amplitude} > 1.2\log_{10}(\mathrm{Period}/\mathrm{days})-2.22$; $G_\mathrm{BP}-G_\mathrm{RP} > 7\log_{10}(\mathrm{Period}/\mathrm{days})-13.20$. The resulting number of O-rich Mira variable candidates was $46\,107$.

\subsection{Assigning distances}
The distance modulus, $m$, of O-rich Mira stars are estimated from the period--luminosity relation
\begin{equation}
    M_{KJK} = \left\{
    \begin{array}{rcl}
       -7.53 - 4.05(\log_{10}P-2.3),  & & \log_{10}P < 2.6,  \\
       -8.75 - 6.99(\log_{10}P-2.6) , & & \log_{10}P \geq 2.6,
    \end{array}
    \right.
    \label{plr}
\end{equation}
where $P$ is the period in days and $M_{KJK}$ the absolute Wesenheit magnitude, and the corresponding apparent Wesenheit magnitude $W_{KJK}$ is
\begin{equation}
    W_{KJK} = K_s - 0.473(J-K_s).
\end{equation}
The extinction coefficient is taken from \cite{WangChen2019}. 
This extinction coefficient does not include the reddening caused by the circumstellar dust if its properties are different from the interstellar dust. Instead, because the period-luminosity relation is calibrated with respect to the O-rich Mira variables in the LMC, the reddening from circumstellar dust has already been considered in equation~\eqref{plr}. The only caveat left is the potential difference in properties of the circumstellar dust between O-rich Mira variables in the LMC and the Milky Way possibly arising due to the difference in metallicity. We consider this a minor effect in our analysis, particularly at shorter periods where significant circumstellar dust is uncommon \citep{Ita2011}.

The intrinsic scatter $\sigma$ of the period-luminosity relation is
\begin{equation}
    \sigma =  \left\{
    \begin{array}{rcl}
       \sigma_{23} + m_{\sigma_1} (\log_{10}P -2.3),  & & \log_{10}P < 2.6,  \\
       \sigma_{23} + 0.3 m_{\sigma_1} + m_{\sigma_2} (\log_{10}P-2.6),  & & \log_{10}P \geq 2.6,
    \end{array}
    \right.
\label{eqn::scatter_plr}
\end{equation}
where $\ln \sigma_{23} = -1.47$, $m_{\sigma_1} = 0.20$ and $m_{\sigma_2} = 0.89$. These relationships are taken from fits of the single-epoch 2MASS data for Mira variables in the LMC (Sanders, in prep.). The scatter is a combination of the single-epoch scatter and the intrinsic scatter due to variance in the population. \cite{Whitelock2008} has argued from a comparison of LMC Mira variables with local Mira variables with Hipparcos and VLBI parallaxes that the Mira variable period-luminosity relation is metallicity-independent, validating our use of the LMC relations for the Milky Way disc Mira variables. Sanders (in prep.) has shown that the $W_{KJK}$ relations for the Milky Way are quite similar to the LMC relations. To compute the uncertainties in distance modulus, $\sigma_m$, we combine in quadrature the intrinsic scatter of the period--luminosity relation from equation~\eqref{eqn::scatter_plr} with the uncertainty propagated from the photometric and period measurement uncertainties. The typical period uncertainties give rise to a median scatter of $0.06\,\mathrm{mag}$ but the scatter arising from the single-epoch measurements is $\gtrsim0.22\,\mathrm{mag}$. Note that the period uncertainties are only meaningful if the correct periodogram peak has been identified. In the case of aliases, the reported period can be formally inconsistent with the true period. \cite{Lebzelter2022} show the impact of aliasing is low. Additionally, in our modelling, we allow for the possibility of a star to be an `outlier' which will capture any incorrectly assigned periods.

\subsection{Gaia astrometric data quality}
LPV stars are one of the most challenging regimes for the Gaia astrometric pipeline for a number of reasons. First, these sources are very red and Gaia's image parameter determination is not well characterised for sources redder than $\nu_\mathrm{eff}=1.24\,\mu\mathrm{m}^{-1}$ \citep{Rowell2021}. Secondly, LPVs are variable whilst the current Gaia astrometric pipelines utilise a fixed colour in the modelling that could lead to systematics \citep{Pourbaix2003}. Finally and possibly most importantly, LPVs can have radii of $1\,\mathrm{AU}$ or larger, and in the optical the photocentres wobble on the order of $\lesssim10\percent$ the radius of the star \cite{Chiavassa2011,Chiavassa2018}. This additional photocentre wobble can lead to biases in the recovered astrometry \citep[e.g.][]{Andriantsaralaza2022} but as the motion is somewhat random and importantly not aligned in any special directions with respect to the parallactic and proper motion directions, particularly when averaging over many stars, the predominant effect is that the reported astrometric uncertainties are underestimates of the true uncertainties.

Sanders (in prep.) has looked at the expected performance of Gaia on a set of modelled Mira variable stars and found that the parallax uncertainties must be inflated for higher parallax objects. This analysis agreed approximately with a full characterisation of the period--luminosity relation and Gaia parallaxes for the Mira variable stars for which Sanders (in prep.) measured an inflation factor of $1+\exp[-(m-8.5)/0.8]$ for the parallax uncertainties. Here, we assume that the proper motion uncertainties must be inflated by the same factor (as validated by Sanders, in prep.). We do not consider the parallaxes in this work.

In addition to the inflation of the astrometric uncertainties on purely physical grounds, any mischaracterisation of Gaia's performance gives rise to misestimated astrometric uncertainties. Steps are taken to mitigate against this in the Gaia pipeline \citep{Lindegren2012} but several studies have shown that problems likely still exist \citep[e.g.][]{ElBadryRix2021,MaizApellaniz2021}. Again, this is particularly a concern for the redder sources due to the image parameter determination. Sanders (in prep.) has modelled the period--luminosity relation using the Gaia parallaxes including a flexible model for the factor by which Gaia's parallax errors must be inflated. The model is two quadratics in $G$ and $\nu_\mathrm{eff}$ for the $5-$ or $6-$parameter astrometric solutions respectively. We adopt their models for the $W_{KJK}$ period--luminosity fits which typically require the parallax uncertainties to be inflated by a factor $\sim1.5$. Although the inflation factor is appropriate for parallax errors, the astrometric modelling is a linear regression so underestimates in the output parameters reflect misestimates of the individual epoch astrometric (along-scan) measurements. It is therefore appropriate to assume all the astrometric uncertainties must be scaled in a similar way to the parallax uncertainties.

\subsection{Final spatial cuts}
We adopt a final series of spatial cuts to focus on Galactic disc members. We remove stars with $270^\circ<\ell<290^\circ$, $-42^\circ<b<-22^\circ$, and $40<\mathrm{heliocentric\ distance}\,(\mathrm{kpc})<60$ to remove potential LMC candidates. As we only consider Mira variables from the Galactic disc, we removed possible bar-bulge contribution by cutting stars with $R<5\, \mathrm{kpc}$, where $R$ is the galactocentric radius. For the interest of kinematic modelling, we only looked at stars with $\mathrm{heliocentric\ distance}<8\, \mathrm{kpc}$ and $R<10\, \mathrm{kpc}$. Stars with $\sigma_m > 0.6$ are removed to avoid stars with extremely large spatial uncertainties. With all of the cuts described in this section, there remain $8\,290$ O-rich Mira variable star candidates in the sample.

\section{Kinematic modelling using dynamical models}\label{sec::model}
Due primarily to the specifics of the scanning law, Gaia's detection of variable stars is a strong function of on-sky location and magnitude. This makes fitting density, or full dynamical, models to any Gaia variable dataset difficult without a careful characterisation of the selection function. Here we employ a simpler approach by only considering the velocity, $\bs{v}$, distribution of our sample at each observed Galactic location, $\bs{x}$ i.e. $p(\bs{v}|\bs{x})$. Except in the most extreme cases, a Mira variable star will not fail to be in the catalogue as a result of its proper motion such that we can safely model the conditional distribution of the proper motions given position. We opt to work with full dynamical models $f(\bs{J})$ expressed as functions of the actions $\bs{J}$ due to their ability to capture the detailed shapes of the velocity distributions and their necessary linking of the radial and azimuthal velocity profiles.

\begin{figure}
    \centering
    \includegraphics[width = \columnwidth]{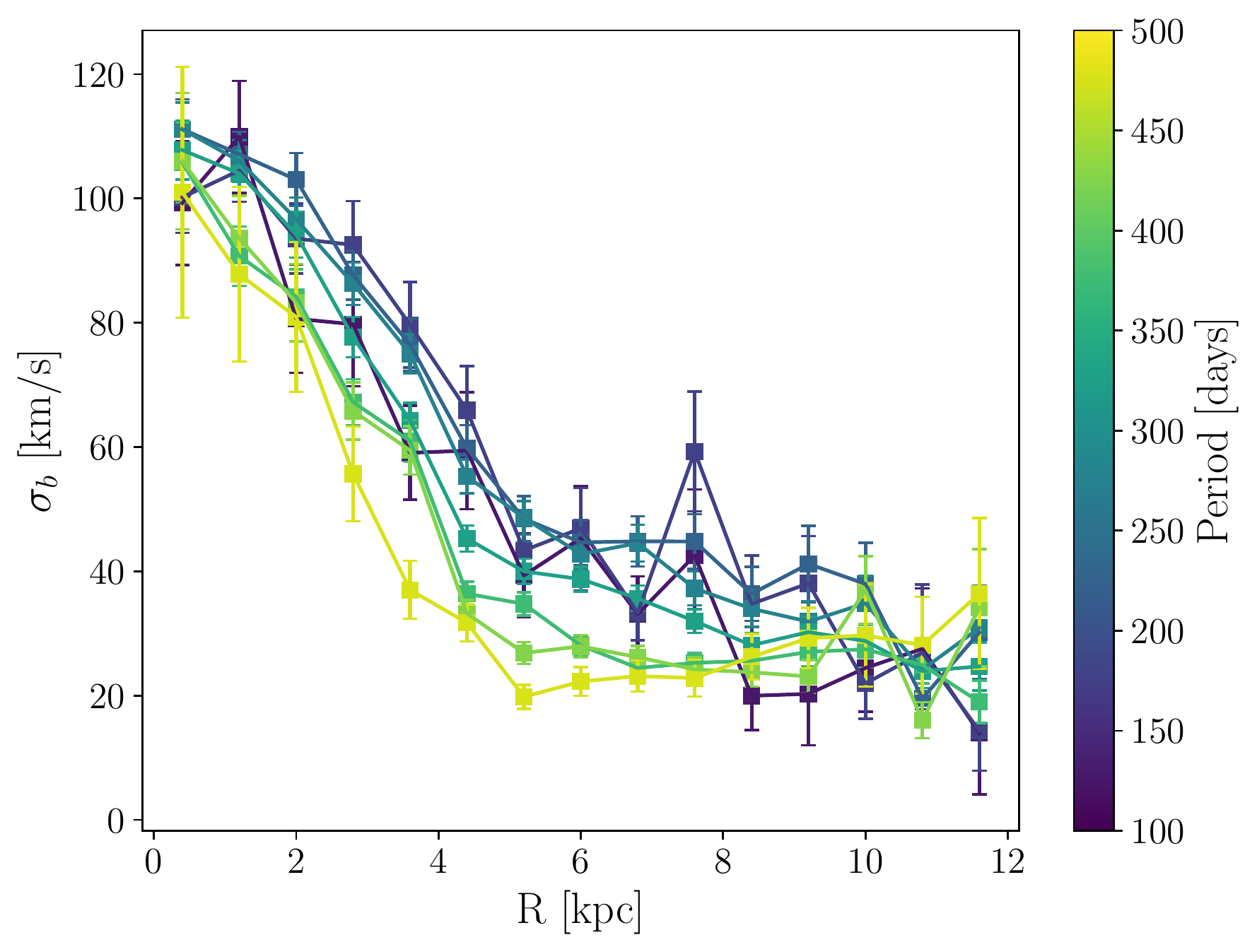}
    \caption{The transverse latitudinal velocity, $v_b$, dispersion profiles of O-rich Mira separated into different period bins. Stars in this figure are only from $|b|<5^{\circ}$, so $v_b$ is approximately equal to the Galactic vertical velocity, or $v_z$, dispersion.}
    \label{vb_dispersion_profile}
\end{figure}

In Fig.~\ref{vb_dispersion_profile}, we plot the latitudinal velocity dispersion profile for several period bins of the selected O-rich Mira as shown. A clear trend in period--dispersion relation is seen implying that the O-rich Mira variables follow a period--age relationship. In our modelling procedure, we will model populations of stars in period bins. Note that the periods are uncertain (as described in the previous section), but typically the uncertainty in the period is small ($\sim10\,\mathrm{day}$, except in the case of aliases) and mixing between bins is a small effect. Working with binned data significantly simplifies our procedure and allows us to fully explore the kinematics with period rather than imposing some functional form. We discuss this latter possibility later.

For a given population of stars with similar periods, we wish to fit the probability distribution function $p(\bs{\mu}|\ell,b,m)$ where $\bs{\mu}$ is the proper motion vector, $(\ell,b)$ the Galactic coordinates and $m$ the distance modulus (as described in the previous section). We begin by writing
\begin{equation}
    p(\bs{\mu}|\ell,b,m) = \frac{p(\ell,b,m,\bs{\mu})}{p(\ell,b,m)}
    = \frac{\int\mathrm{d}v_{||}p(\ell,b,m,\bs{\mu},v_{||})}{\int\mathrm{d}^2\bs{\mu}\,\mathrm{d}v_{||}\,p(\ell,b,m,\bs{\mu},v_{||})}.
\label{eqn}
\end{equation}
The proper motions and distance moduli are measured quantities with some associated uncertainties characterised by the proper motion covariance matrix $\bs{\Sigma}_\mu$ and the uncertainty in distance modulus $\sigma_m$. We, therefore, marginalize over the uncertainties by writing
\begin{equation}
\begin{split}
    p(\ell,&b,m,\bs{\mu},v_{||}) \\&= \int\mathrm{d}^2\bs{\mu}'\mathrm{d}m'\mathcal{N}(\bs{\mu}|\bs{\mu}',\bs{\Sigma}_\mu)\mathcal{N}(m|m',\sigma^2_m)p(\ell,b,m',\bs{\mu}',v_{||}),
\end{split}
\label{p_marginalise}
\end{equation}
where $\mathcal{N}(x|\mu,\sigma^2)$ are Gaussians with mean $\mu$ and variance $\sigma^2$. We then relate the distribution in observable coordinates to the dynamical distribution function in actions as
\begin{equation}
    p(\ell,b,m',\bs{\mu}',v_{||}) = \left|\frac{\partial (\bs{J}, \bs{\theta})}{\partial (\ell,b, m, \bs{\mu}, v_{||})}\right|f(\bs{J}) \propto  s^5\cos b\,f(\bs{J}),
\end{equation}
where $\bs{J}=(J_r,J_\phi,J_z)$ is the set of actions corresponding to the observed 6d coordinate (with corresponding angle coordinates $\bs{\theta}$) and $s$ is the distance corresponding to distance modulus $m$. Note the Jacobian between $(\bs{x},\bs{v})$ and $(\bs{J},\bs{\theta})$ is unity due to the canonical nature of the action-angle coordinates.

We choose $f(\bs{J})$ as a quasi-isothermal distribution function, which is suitable for warm discs \citep{Binney2010}. We follow the implementation in \textsc{Agama} \citep{Agama} which has a functional form given by
\begin{align}
&f(\bs{J}) = \frac{\tilde\Sigma\,\Omega}{2\pi^2\,\kappa^2} \times
\frac{\kappa}{\tilde\sigma_r^2} \exp\left(-\frac{\kappa\,J_r}{\tilde\sigma_r^2}\right) \times
\frac{\nu}   {\tilde\sigma_z^2} \exp\left(-\frac{\nu\,   J_z}{\tilde\sigma_z^2}\right) \times B(J_\phi),\nonumber\\
&B(J_\phi)=\left\{ \begin{array}{ll}  1 & \mbox{if }J_\phi\ge 0, \\
\exp\left( \frac{2\Omega\,J_\phi}{\tilde\sigma_r^2} \right) & \mbox{if }J_\phi<0, \end{array} \right.,\nonumber\\
&\tilde\Sigma(R_\mathrm{c})  \equiv \Sigma_0 \exp( -R_\mathrm{c} / R_\mathrm{disc} ) ,\nonumber\\
&\tilde\sigma_r^2(R_\mathrm{c}) \equiv \sigma_{r,0}^2 \exp( -2(R_\mathrm{c}-R_0) / R_{\sigma,r} ),\nonumber\\
&
\tilde\sigma_z^2(R_c) \equiv \sigma_{z,0}^2 \exp( -2(R_c-R_0) / R_{\sigma,z} ),
\end{align}
where $R_\mathrm{c}$ is the radius corresponding to a circular orbit of angular momentum $J_\phi\equiv L_z$ and $(\kappa,\Omega,\nu)$ are the epicyclic frequencies at this angular momentum.
This distribution function describes an approximately exponential disc in radius which is broadened/warmed vertically and radially by two exponential terms. There are five key free parameters for the model: (i) the scalelength of the disc, $R_\mathrm{disc}$, (ii) the radial ($\sigma_{r,0}$) and vertical ($\sigma_{z,0}$) normalizations of the velocity dispersions at the Sun ($R=R_0$), and (iii) their corresponding scalelengths ($R_{\sigma,r}$ and $R_{\sigma,z}$). The actions are evaluated using the `St\"ackel fudge' algorithm described by \cite{Binney2012}, summarized and critically assessed against alternatives in \cite{SandersBinney2016} and implemented in \textsc{Agama} \citep{Agama}. We adopt a fixed axisymmetric gravitational potential for the Galaxy from \cite{McMillan2017}. Fixing the potential could lead to sub-optimal model fits (as we will discuss later) but it significantly simplifies the computation and incorporates external constraints from the analysis of other datasets.

\subsection{Computational considerations}\label{section::computation}
The computational difficulty in evaluating equation~\eqref{eqn} is computing the integrals efficiently. Here we use Monte Carlo integration. For the numerator, we generate a set of $N$ samples for each star from the proper motion and distance modulus error ellipses. The unknown $v_{||}$ is sampled from a probability distribution $G(v_{||}|\ell, b, m, \bs{\mu})$ which is proportional to a quasi-isothermal distribution function with fixed parameters $f'(\bs{J})$ at a given $(\ell, b, m, \bs{\mu})$,
\begin{equation}
    G(v_{||}|\ell, b, m, \bs{\mu})= \frac{p(\ell,b,m,\bs{\mu},v_{||})}{\int\mathrm{d}v_{||}\,p(\ell,b,m,\bs{\mu},v_{||})} = A_{v_{||}}f'(\bs{J}).
    \label{G_v||}
\end{equation}
Samples are generated from this distribution using the inverse cumulative distribution.
The value of $f'(\bs{J}_i)$ for each sample is stored to reweight the Monte Carlo sum. For the denominator, we sample $\bs{v}=(v_{x},v_{y},v_{z})$ directly at a given observed position $(\ell, b, m)$ in a similar way to the numerator as
\begin{equation}
    G(\bs{v}|\ell, b, m)= \frac{p(\ell,b,m, \bs{v})}{\int\mathrm{d^3}\bs{v}\,p(\ell,b,m,\bs{v})} =  A_{\bs{v}}f'(\bs{J}),
    \label{G_v3}
\end{equation}
from which samples are generated using Markov Chain Monte Carlo \citep[MCMC, ][]{emcee}, and once again $f'(\bs{J}_i)$ are stored. $A_{v_{||}}$ and $A_{\bs{v}}$ defined in equation~\eqref{G_v||} and~\eqref{G_v3} are constant factors which can be computed for each individual star. Only the ratio of these two normalisation factors is important:
\begin{equation}
    A \equiv\frac{A_{\bs{v}}}{A_{v_{||}}}= \frac{\int\mathrm{d}v_{||}\,p(\ell,b,m,\bs{\mu},v_{||})}{\int\mathrm{d^3}\bs{v}\,p(\ell,b,m,\bs{v})} = \frac{\int\mathrm{d}v_{||}\, f'(\bs{J})}{\int\mathrm{d^3}\bs{v}\,f'(\bs{J})}.
    \label{normalisation_constant}
\end{equation}
$A$ is evaluated using Monte Carlo integration: $v_{||}$ and $\bs{v}$ are sampled from a Gaussian distribution centred on zero in the radial and vertical velocities, and on the rotation curve in the azimuthal velocity. As $f'(\bs{J})$ is fixed, $A$ can be precomputed once for each individual star to a desired accuracy.
The $f'(\bs{J})$ we use throughout this paper has fixed parameters: $R_\mathrm{disc} = 2.5\,\mathrm{kpc}$, $\sigma_{r,0} = 50\,\mathrm{km/s}$, $\sigma_{z,0} = 50\,\mathrm{km/s}$, $R_{\sigma,r} = 5.0\,\mathrm{kpc}$ and $R_{\sigma,z} = 5.0\,\mathrm{kpc}$. These parameters are chosen such that the distributions of the integration samples are typically broader than the modelled distributions to minimise bias in the Monte Carlo integration. Sampling from the distribution $G$, instead of a Gaussian distribution increases the computational efficiency by reducing the noise in the Monte Carlo integration for a fixed number of sampling. Now for each star, the integrals (up to a normalization constant) are given by

\begin{equation}
    p(\ell,b,m,\bs{\mu}) \approx \frac{1}{N A_{v_{||}}}\color{black}\sum^{\substack{\mathrm{errors\,in\,}m, \bs{\mu}\\v_{||}\mathrm{\,from\,}G(v_{||}|\dots)}}_i s_i^5\cos b \frac{f(\bs{J}_i)}{f'(\bs{J}_i)},
    \label{num}
\end{equation}
and
\begin{equation}
    p(\ell,b,m) \approx \frac{1}{NA_{\bs{v}}}\color{black}\sum^{\substack{\mathrm{errors\,in\,}m\\\bs{v}\mathrm{\,from\,}G(\bs{v}|\dots)}}_i s_i^3\cos b \frac{f(\bs{J}_i)}{f'(\bs{J}_i)}.
    \label{denom}
\end{equation}
Note in the second expression we only have $3$ powers of $s$ as the integral has been rewritten in terms of the 3d space velocity $\bs{v}$ (as opposed to the observable space of proper motion and radial velocity). As we are using a fixed potential, we precompute $\bs{J}_i$, $R_{\mathrm{c},i}$ and the epicyclic frequencies for all samples using the routines from \cite{Agama} and \cite{Bovy2015}.

\subsection{Outlier component}\label{section::outlier_compt}
Another complexity is to introduce an outlier distribution to overcome the contamination of samples by stars which are members of the halo, are possibly not Mira variable stars or have poorly-measured periods. To do this, we assume that the velocity distribution of the contamination stars is described by a 3D spherically symmetric Gaussian distribution that is centred on Galactocentric $\bs{v} = \bs{0}$ with standard deviation in each dimension $\sigma_v$. Similar to the previous approach, we calculate the $p_{\mathrm{outlier}} (\bs{\mu}|\ell,b,m)$ using equation~\eqref{eqn} and \eqref{p_marginalise}, but replacing $p(\ell,b,m',\bs{\mu}',v_{||})$ with $p_{\mathrm{outlier}}(\ell,b,m',\bs{\mu}',v_{||})$ which is chosen to be
\begin{equation}
    p_{\mathrm{outlier}}(\ell,b,m',\bs{\mu}',v_{||}) = s^5 \cos b\,\mathcal{N}(\bs{v}|\bs{0},\sigma_{{v}}^2\bs{I}) \mathcal{U}(x,y,z),
\end{equation}
where $\mathcal{U}(x,y,z)$ is the uniform distribution in Galactocentric Cartesian spatial coordinates $(x,y,z)$. For each star, $p_{\mathrm{outlier}} (\bs{\mu}|\ell,b,m)$ is evaluated numerically by
\begin{equation}
    p_{\mathrm{outlier}} (\bs{\mu}|\ell,b,m) = \frac{\sum^{\mathrm{errors\,in\,}m, \bs{\mu}}_i s_i^5 \mathcal{N}(s_i\bs{\mu}_i+\bs{v}_{t,\odot,i}|\bs{0},\sigma^2_v\bs{I})}{\sum^{\mathrm{errors\,in\,}m}_i s_i^3},
\end{equation}
where $v_{t,\odot}$ is the solar velocity in the Galactocentric frame projected in the plane perpendicular to the line-of-sight between the star and the Sun. To include this distribution in the log-likelihood, we rewrite the probability for each individual star as
\begin{equation}
    p_{\mathrm{tot}, j} = (1-\epsilon)p_{j}(\bs{\mu}|\ell,b,m)+\epsilon p_{\mathrm{outlier}, j}(\bs{\mu}|\ell,b,m).
    \label{P_outlier}
\end{equation}
Note with this definition, the outlier fraction at each spatial location, $\epsilon$, is approximately constant. 
We choose the Gaussian because the contamination could come from a variety of sources, and the Gaussian distribution is a general, easily-computed way to characterise those sources.

\subsection{Likelihood}\label{section::logL}
We have now fully specified our model. The full log-likelihood for each population of stars is
\begin{equation}
   \ln L = \sum^{\mathrm{stars}}_j \ln p_\mathrm{tot}(\bs{\mu}_j|\ell_j,b_j,m_j),
   \label{log-likelihood_full}
\end{equation}
For each population of stars, we optimize the likelihood with respect to the five parameters of the quasi-isothermal ($R_\mathrm{disc}$, $\sigma_{r,0}$, $\sigma_{z,0}$, $R_{\sigma,r}$, $R_{\sigma,z}$) and the two parameters of the outlier distribution $(\epsilon, \sigma_v)$. The log-likelihood is explored using MCMC performed with \textsc{emcee} \citep{emcee}. We adopt priors on the radial scale lengths as $R_\mathrm{disc}\sim\mathcal{N}(3.8\,\mathrm{kpc}, (2\,\mathrm{kpc})^2)$ and $R_{\sigma,r/z}\sim\mathcal{N}(4.5\,\mathrm{kpc}, (3\,\mathrm{kpc})^2)$ and a prior for velocity dispersion of the outlier component is a normal distribution $\sigma_{\bs{v}}\sim\mathcal{N}(200\,\mathrm{km\,s}^{-1}, (150\,\mathrm{km\,s}^{-1})^2)$. The priors for the other three parameters are uniform: $\sigma_{r/z, 0}\sim\mathcal{U}(0,120\,\mathrm{km\,s}^{-1})$ and $\epsilon\sim\mathcal{U}(0,1)$.

A final step in our procedure is converting the modelled distribution function parameters to the physical measures of the velocity dispersion in the solar neighbourhood. It is these quantities we compare with previous characterisations of the age--velocity dispersion relation. For each set of ($R_{\mathrm{disc}}$, $\sigma_{r,0}$, $\sigma_{z,0}$, $R_{\sigma,r}$, $R_{\sigma,z}$), we generate mock stars using the \textsc{Agama} DF sampling routines and fit an exponential profile $\sigma_{i} = \widetilde{\sigma}_{i,0}\exp[{(R_0-R)/\widetilde{R}_{\sigma,i}}]$ to the radial and vertical velocity dispersions binned in radius.
The normalization $\widetilde{\sigma}_{i,0}$ and scalelength $\widetilde{R}_{\sigma,i}$ give the physical velocity dispersion and its radial gradient in the solar neighbourhood. 

\subsection{Mock samples and validation}\label{sec:mock}
\begin{figure*}
    \centering
    \includegraphics[width = \textwidth]{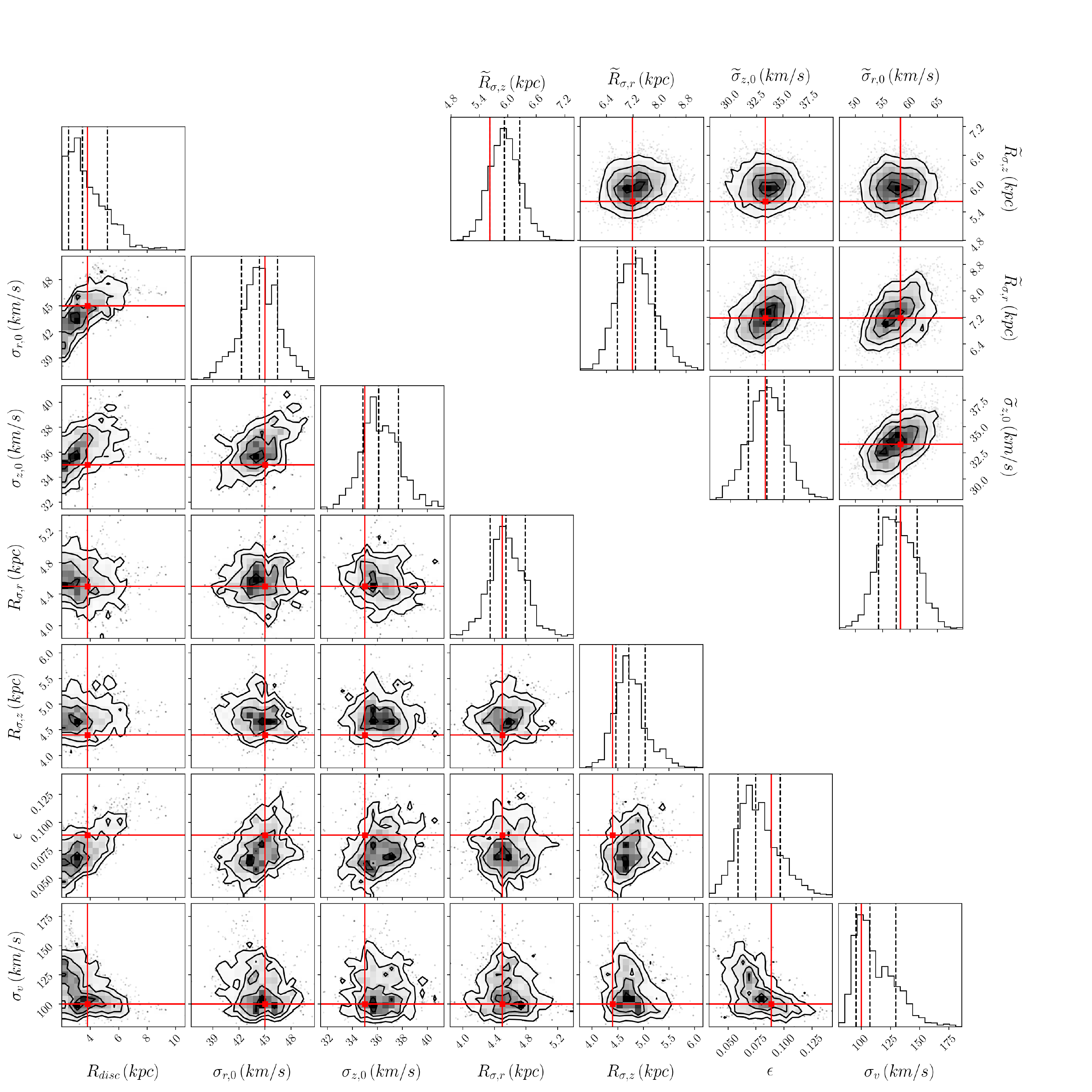}
    \caption{Results of fits on mock data: the lower left corner plot is the posterior of the fitting parameters from the test on mock data including an outlier distribution. The red lines are the parameters that generated the mock sample, and the black dashed lines are the $16$th, $50$th and $86$th percentiles of the posterior, respectively. The upper right corner plot gives the posteriors of the physical velocity dispersion parameters corresponding to the sets of fitted parameters. The physical velocity dispersion parameters are propagated from the fitted parameters using the routine described in Section~\ref{sec::application}.}
    \label{mock_posterior_outlier}
\end{figure*}
Given a fitted $f(\bs{J})$ model, we wish to draw mock samples to compare with the data and validate our fitting procedure. We use the \textsc{Agama} DF sampling routine to generate a large number of mock stars. For each generated mock star, we find the nearest observed star in our dataset in $(R,z)$, place the mock star at the azimuth $\phi$ of the real star and transform the mock polar velocities to $\bs{\mu}$. This procedure exploits the axisymmetry of the models. We further scatter the proper motions and distance moduli of the mock stars by the corresponding uncertainties of the real stars. The previous Mira selection criteria in the heliocentric distance and $R$ are also applied to the mock sample. Note this procedure produces a mock dataset with each real star corresponding to multiple mock stars in proportion to the local stellar density at the location of the real star. This reduces the shot noise in our mock samples but means the mock sample has a different spatial density to the data. To reproduce the spatial distribution of the dataset, we record the index of the closest matched real star for each mock star and then count the number of times that this real star is the closest match to any mock star. A weight is calculated for each mock star as the reciprocal of this number count. The weight will be used when we compare our fitted model to the dataset. When directly comparing to a fitted dataset, we further remove mock stars which do not reside within $100 \, \mathrm{pc}$ of any real star (this requirement is not imposed on the mock test set described below but makes little practical difference). Our procedure does not fully generate the data as we have not accounted for uncertainty in the data $(R,z)$. However, it is sufficient for validation purposes.

We can use the generated mock observations to test the validity of our method. We generate a mock sample of $614$ stars from $f(\bs{J})$ with known parameters chosen arbitrarily as $R_\mathrm{disc} = 3.8\,\mathrm{kpc}$, $\sigma_{r,0} = 45.0\,\mathrm{km\,s}^{-1}$, $\sigma_{z,0} = 35.0\,\mathrm{km\,s}^{-1}$, $R_{\sigma,r} = 4.5\,\mathrm{kpc}$ and $R_{\sigma,z} = 4.4\,\mathrm{kpc}$. We then replace velocities of $10\%$ of the generated data with $\bs{v}$ sampled from a spherically symmetric Gaussian $\mathcal{N}(\bs{v}|\bs{0}, (100 \, \mathrm{km/s})^2\bs{I})$ which is the assumed velocity distribution of outlier stars. Stars sampled from the outlier distribution have a chance to be unbound from the potential, so after removing those unbound stars, the actual proportion of outlier stars can be smaller than $10\%$, i.e. $\epsilon < 10\%$. Without those high-velocity stars in the mock sample, the velocity dispersion of the generated outlier stars is reduced so a fitted $\sigma_{{v}}<100 \, \mathrm{km/s}$ is expected but the recovered parameters of the $f(\bs{J})$ model should be unbiased.
The posteriors from the MCMC are shown in the low left of Fig.~\ref{mock_posterior_outlier}.
The parameters $\sigma_{z,0}$ and $R_{\sigma_{z, 0}}$ both deviate slightly from the default parameters but only around the 1$\sigma$ level. In the upper right corner of Fig.~\ref{mock_posterior_outlier}, we convert each set of fitted parameters into the physical velocity dispersion profile parameters, $\widetilde{\sigma}_{i,0}$ and $\widetilde{R}_{\sigma,i}$. Although there are small differences in the distribution function parameters, the resulting physical velocity dispersions and scalelengths at the solar position are well recovered. We also produced the posterior of the same sample using the log-likelihood without the outlier distribution. The medians of the parameters are $R_\mathrm{disc} = 3.56\,\mathrm{kpc}$, $\sigma_{r,0} = 65.74\,\mathrm{km\,s}^{-1}$, $\sigma_{z,0} = 43.16\,\mathrm{km\,s}^{-1}$, $R_{\sigma,r} = 3.14\,\mathrm{kpc}$ and $R_{\sigma,z} = 2.48\,\mathrm{kpc}$. As expected, $\sigma_{r,0}$ and $\sigma_{z,0}$ are overestimated. This demonstrates that adding the outlier distribution is necessary when the contamination of the sample is significant.

\section{Velocity dispersion of O-rich Mira variable stars in different age bins}\label{sec::application}
\begin{figure*}
    \centering
    \includegraphics[width=\textwidth]{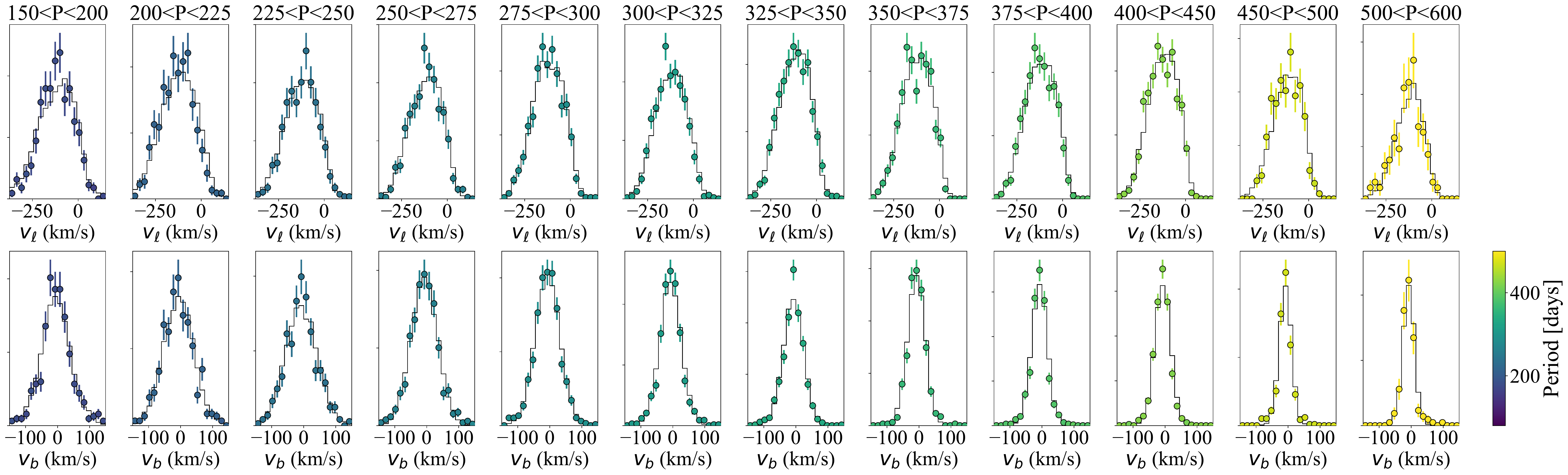}
    \caption{
    Velocity histograms for O-rich Mira variables separated by period (as given in days above each column). The top panels show $v_\ell$ and bottom $v_b$. The points are data and black lines the models.}
    \label{v_l_v_b_distribution}
\end{figure*}

\begin{table*}
\caption{Distribution function parameter estimates for the Mira variable model fits. The left column gives the considered period bin and the other columns show the median and uncertainties estimated from the $16$th and $84$th percentiles.}
    \begin{center}
        \begin{tabular}{cccccccccc}
 \shortstack{Period\\range (day)}  &  \shortstack{Mean\\period (day)} &  \shortstack{Number\\of stars}  & $R_{\mathrm{disc}}\,\mathrm{(kpc)}$       & $\sigma_{r,0}\,\mathrm{(km/s)}$   & $\sigma_{z,0}\,\mathrm{(km/s)}$   & $R_{\sigma,r}\,\mathrm{(kpc)}$   & $R_{\sigma,z}\,\mathrm{(kpc)}$  &  $\epsilon$             & $\sigma_{\bs{v}}\,\mathrm{(km/s)}$                \\
\hline
 $80-150$          &     126.4     &       230        & $3.55^{+2.40}_{-1.19}$ & $48.05^{+6.17}_{-4.83}$ & $30.45^{+2.14}_{-2.01}$ & $8.32^{+3.29}_{-1.64}$  & $9.65^{+2.28}_{-2.41}$  & $0.06^{+0.02}_{-0.02}$ & $151.49^{+30.72}_{-22.91}$ \\
 $150-200$         &     179.3     &       430        & $3.56^{+1.27}_{-0.74}$ & $40.59^{+3.06}_{-3.26}$ & $41.07^{+3.62}_{-3.46}$ & $3.54^{+0.25}_{-0.19}$  & $6.47^{+1.14}_{-1.10}$  & $0.02^{+0.02}_{-0.02}$ & $105.58^{+61.84}_{-40.22}$ \\
 $200-225$         &     212.8     &       442        & $5.11^{+1.00}_{-1.35}$ & $53.14^{+5.51}_{-5.39}$ & $51.72^{+3.39}_{-2.89}$ & $5.05^{+1.19}_{-0.60}$  & $9.52^{+2.06}_{-1.94}$  & $0.01^{+0.04}_{-0.01}$ & $46.46^{+111.19}_{-21.72}$ \\
 $225-250$         &     237.7     &       494        & $3.79^{+1.43}_{-0.82}$ & $37.66^{+2.66}_{-2.18}$ & $55.99^{+3.74}_{-4.19}$ & $3.97^{+0.32}_{-0.31}$  & $10.36^{+2.89}_{-2.19}$ & $0.01^{+0.01}_{-0.01}$ & $78.35^{+71.38}_{-36.07}$  \\
 $250-275$         &     263.3     &       708        & $3.53^{+1.95}_{-0.91}$ & $52.38^{+3.59}_{-3.72}$ & $42.22^{+1.83}_{-2.10}$ & $12.16^{+3.37}_{-2.72}$ & $7.41^{+1.30}_{-0.92}$  & $0.03^{+0.02}_{-0.01}$ & $79.23^{+28.35}_{-12.22}$  \\
 $275-300$         &     287.2     &       909        & $4.47^{+1.55}_{-0.98}$ & $51.79^{+2.49}_{-2.11}$ & $39.57^{+2.00}_{-1.75}$ & $13.57^{+3.08}_{-2.17}$ & $7.24^{+1.27}_{-0.87}$  & $0.02^{+0.01}_{-0.01}$ & $104.36^{+22.33}_{-15.63}$ \\
 $300-325$         &      313.0      &       907        & $2.72^{+0.71}_{-0.58}$ & $46.46^{+2.04}_{-2.18}$ & $34.53^{+1.90}_{-1.72}$ & $11.68^{+2.13}_{-1.49}$ & $8.12^{+1.50}_{-1.15}$  & $0.00^{+0.01}_{-0.00}$ & $109.87^{+45.84}_{-36.79}$ \\
 $325-350$         &     337.7     &       970        & $2.49^{+0.67}_{-0.47}$ & $43.15^{+1.68}_{-1.75}$ & $32.94^{+1.34}_{-1.18}$ & $12.10^{+2.50}_{-1.54}$ & $9.08^{+1.55}_{-1.17}$  & $0.01^{+0.01}_{-0.00}$ & $114.63^{+37.64}_{-21.27}$ \\
 $350-375$         &     362.3     &       861        & $5.29^{+1.78}_{-1.38}$ & $42.44^{+1.82}_{-1.78}$ & $28.84^{+1.26}_{-1.38}$ & $13.52^{+4.08}_{-2.42}$ & $11.35^{+2.55}_{-2.39}$ & $0.01^{+0.01}_{-0.01}$ & $79.10^{+79.99}_{-45.18}$  \\
 $375-400$         &     387.3     &       784        & $4.69^{+2.07}_{-1.41}$ & $42.33^{+2.26}_{-1.62}$ & $23.89^{+1.54}_{-1.49}$ & $12.01^{+3.08}_{-1.96}$ & $7.89^{+2.12}_{-1.27}$  & $0.01^{+0.01}_{-0.01}$ & $79.19^{+48.55}_{-32.91}$  \\
 $400-450$         &     422.5     &       1015       & $2.87^{+1.40}_{-0.69}$ & $41.45^{+1.81}_{-1.61}$ & $25.77^{+0.86}_{-1.04}$ & $14.43^{+3.11}_{-2.01}$ & $13.69^{+2.48}_{-1.58}$ & $0.00^{+0.01}_{-0.00}$ & $88.41^{+66.37}_{-43.88}$  \\
 $450-500$         &     470.9     &       396        & $3.18^{+2.57}_{-1.28}$ & $37.42^{+1.90}_{-2.27}$ & $19.56^{+1.59}_{-1.71}$ & $13.22^{+2.70}_{-2.64}$ & $15.03^{+4.98}_{-4.03}$ & $0.04^{+0.02}_{-0.02}$ & $82.44^{+21.10}_{-18.52}$  \\
 $500-600$         &     527.5     &       144        & $4.68^{+2.68}_{-2.10}$ & $34.27^{+3.22}_{-2.85}$ & $16.85^{+1.92}_{-2.41}$ & $11.45^{+3.82}_{-3.06}$ & $11.34^{+5.52}_{-3.99}$ & $0.02^{+0.03}_{-0.01}$ & $121.84^{+69.88}_{-44.77}$ \\
        \end{tabular}
    \end{center}
    \label{MCMC_parameters}
\end{table*}

\begin{table*}
\caption{Solar neighbourhood velocity dispersions and local spatial gradients of the velocity dispersions for the Mira variable fits. The age estimations are also provided, where $\tau_{r}$ is the age estimation from the radial velocity dispersion while $\tau_{z}$ is that from the vertical velocity dispersion. }
\begin{center}
\begin{tabular}{ccccccc}
 Period (days)   & $\widetilde{\sigma}_{r,0}\,\mathrm{(km/s)}$   & $\widetilde{\sigma}_{z,0}\,\mathrm{(km/s)}$   & $\widetilde{R}_{\sigma,r}\,\mathrm{(kpc)}$   & $\widetilde{R}_{\sigma,z}\,\mathrm{(kpc)}$   
 & $\tau_{r}\,\mathrm{(Gyr)}$        & $\tau_{z}\,\mathrm{(Gyr)}$       \\
\hline
 $80-150$          & $49.83^{4.39}_{3.78}$   & $24.59^{1.47}_{1.38}$   & $10.54^{3.94}_{2.13}$  & $9.04^{1.64}_{1.64}$   & $8.57^{+1.13}_{-0.98}$  & $6.34^{+0.36}_{-0.34}$ \\
 $150-200$         & $67.20^{4.70}_{4.20}$   & $34.44^{1.86}_{1.43}$   & $6.88^{0.62}_{0.61}$   & $7.46^{0.85}_{0.92}$   & $10.82^{+0.56}_{-0.52}$ & $8.07^{+0.82}_{-0.75}$ \\
 $200-225$         & $62.96^{2.99}_{3.11}$   & $38.24^{1.55}_{1.56}$   & $7.92^{1.47}_{0.91}$   & $9.37^{1.48}_{1.27}$   & $10.41^{+0.45}_{-0.45}$ & $9.34^{+0.64}_{-0.66}$ \\
 $225-250$         & $52.83^{3.00}_{3.20}$   & $40.24^{1.67}_{1.78}$   & $6.16^{0.53}_{0.53}$   & $10.10^{1.66}_{1.42}$  & $9.25^{+0.73}_{-0.89}$  & $9.72^{+0.61}_{-0.67}$ \\
 $250-275$         & $51.90^{2.48}_{2.29}$   & $32.92^{1.14}_{1.16}$   & $15.31^{4.51}_{3.61}$  & $7.78^{0.94}_{0.77}$   & $9.09^{+0.72}_{-0.78}$  & $7.58^{+0.60}_{-0.55}$ \\
 $275-300$         & $50.54^{2.07}_{1.91}$   & $31.08^{1.21}_{1.08}$   & $16.61^{4.20}_{2.97}$  & $7.54^{1.05}_{0.76}$   & $8.75^{+0.78}_{-0.70}$  & $7.25^{+0.52}_{-0.44}$ \\
 $300-325$         & $46.67^{1.73}_{1.72}$   & $27.57^{1.11}_{1.01}$   & $14.35^{2.78}_{2.08}$  & $8.05^{1.15}_{0.87}$   & $7.80^{+0.55}_{-0.57}$  & $6.77^{+0.35}_{-0.35}$ \\
 $325-350$         & $43.27^{1.50}_{1.54}$   & $26.27^{0.86}_{0.88}$   & $14.41^{3.15}_{2.00}$  & $8.66^{1.10}_{0.91}$   & $7.01^{+0.59}_{-0.54}$  & $6.61^{+0.34}_{-0.34}$ \\
 $350-375$         & $41.74^{1.67}_{1.69}$   & $23.11^{0.89}_{0.93}$   & $15.35^{4.79}_{2.96}$  & $10.24^{1.74}_{1.75}$  & $6.66^{+0.52}_{-0.49}$  & $6.20^{+0.33}_{-0.30}$ \\
 $375-400$         & $41.98^{1.84}_{1.47}$   & $20.02^{1.00}_{0.94}$   & $13.73^{3.78}_{2.23}$  & $7.82^{1.65}_{1.12}$   & $6.67^{+0.51}_{-0.49}$  & $5.66^{+0.34}_{-0.32}$ \\
 $400-450$         & $41.08^{1.58}_{1.50}$   & $20.92^{0.71}_{0.76}$   & $16.45^{3.56}_{2.46}$  & $11.61^{1.65}_{1.17}$  & $6.43^{+0.47}_{-0.43}$  & $5.86^{+0.31}_{-0.31}$ \\
 $450-500$         & $37.21^{1.87}_{2.03}$   & $16.46^{1.21}_{1.16}$   & $14.60^{3.38}_{2.91}$  & $12.58^{3.00}_{2.60}$  & $5.52^{+0.57}_{-0.57}$  & $4.60^{+0.61}_{-0.96}$ \\
 $500-600$         & $34.24^{2.99}_{2.78}$   & $14.65^{1.34}_{1.85}$   & $12.44^{4.03}_{3.23}$  & $10.46^{3.60}_{3.11}$  & $4.50^{+0.86}_{-1.12}$  & $3.62^{+0.76}_{-1.05}$ \\
\end{tabular}
\end{center}
\label{actual_params}
\end{table*}
\begin{figure*}
    \centering
    \includegraphics[width=\textwidth]{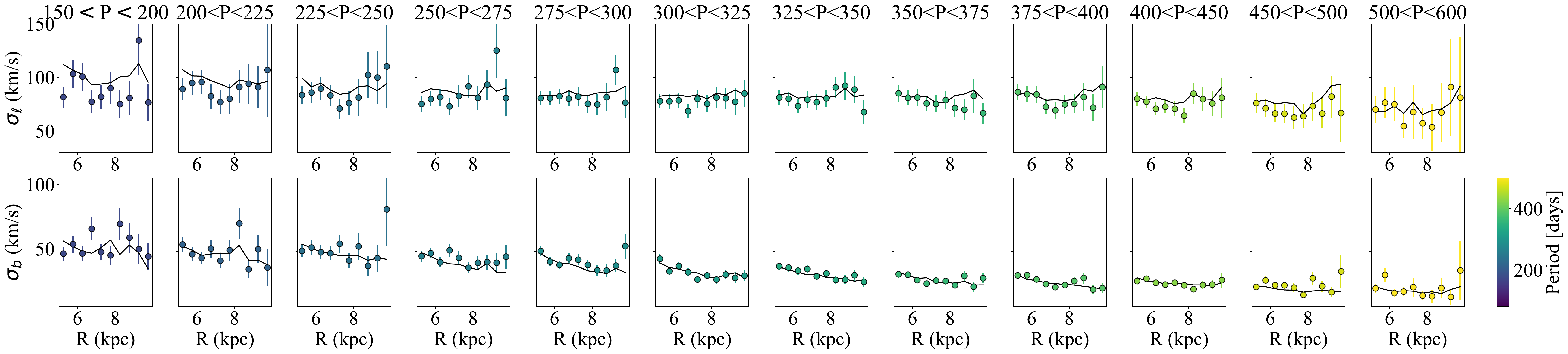}
    \caption{Velocity dispersion profiles as a function of Galactocentric radius for O-rich Mira variables separated by period (as given in days above each column). The top panels show longitudinal, $\ell$, and bottom latitudinal, $b$. The points are data and black lines the models.}
    \label{sigmal_sigmab_radial_profile}
\end{figure*}
\begin{figure*}
    \centering
    \includegraphics[width=\textwidth]{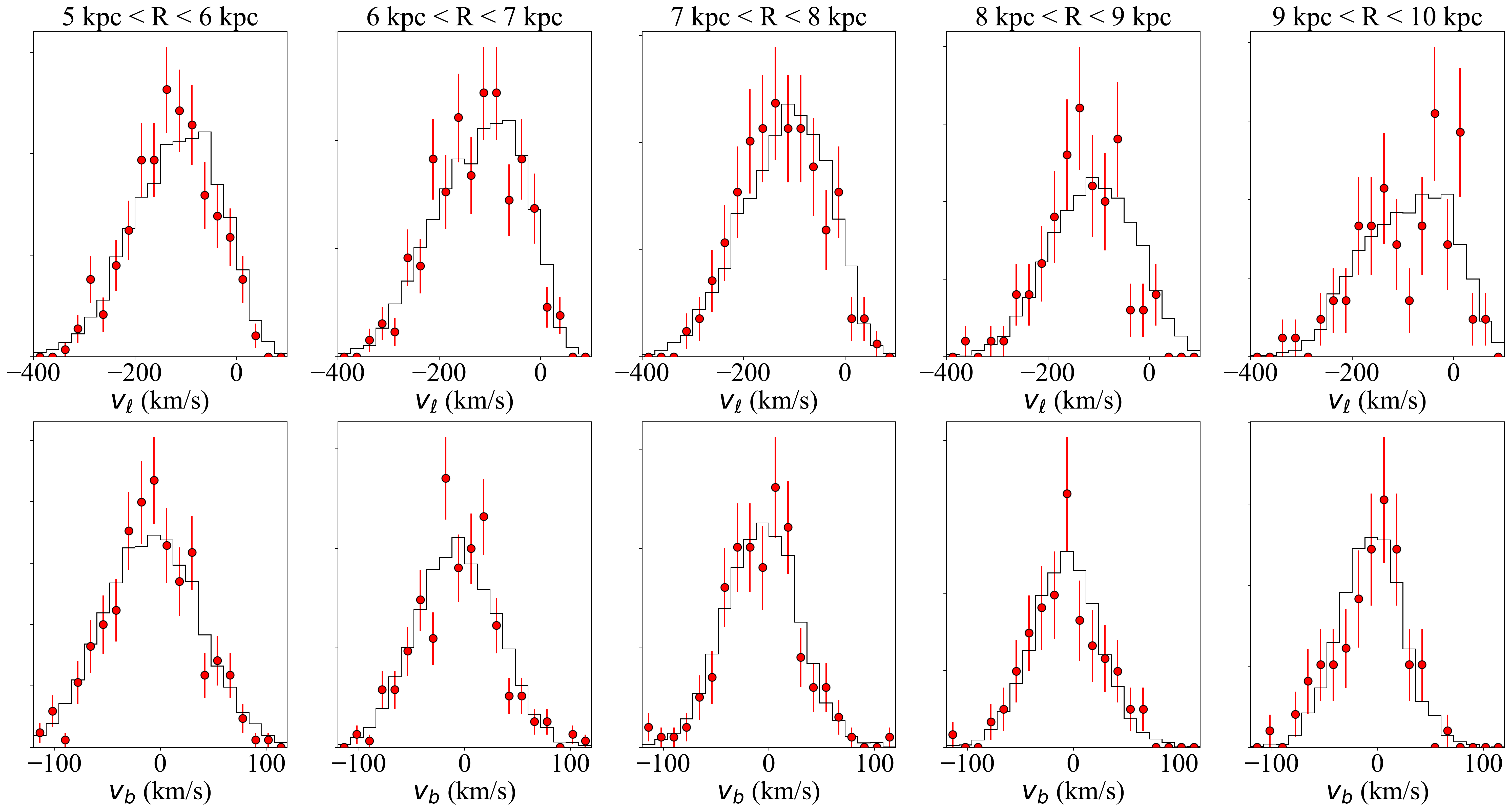}
    \caption{Velocity histograms for O-rich Mira variables with periods in the range $275-300$ day separated into bins of Galactocentric radius (as given above each column). The top panels show the longitudinal velocity $v_\ell$ and the bottom the latitudinal velocity, $v_b$. The red points are data and the black lines are the models.}
    \label{vl_vb_radial_profile}
\end{figure*}
\begin{figure*}
    \centering
    \includegraphics[width=\textwidth]{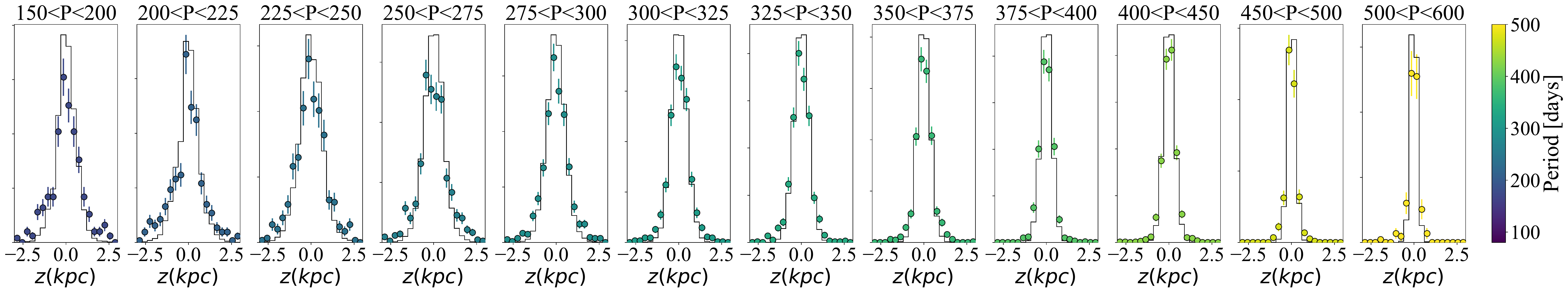}
    \caption{Vertical density distribution profile for O-rich Mira variables separated by period bins (as given in days above each panel). Each panel shows the dataset (points) compared to the unweighted distribution of mock samples (black lines). All histograms are normalised, and subplots do not share the same y-axis. The discrepancy between the distributions is a reflection of the completeness of the dataset.}
    \label{vertical_density_profile}
\end{figure*}
To investigate the kinematic properties of the sample defined in Section~\ref{sec::data}, we put the O-rich Mira variables into period bins and treat stars in each bin as a sub-population drawn from the same DF. We choose the period bins to be wider than the typical uncertainties in the period measurements, and hence we neglect the period uncertainties that scatter stars from bin to bin (the impact of the period uncertainties on the distance uncertainties \emph{have} been considered). The median of the period uncertainties is $11.6$ days and $7.1$ days for those stars with periods less than $300$ days. We have also tried to bin stars with a wider period bin ($50$ days instead of $25$), which gives very similar results to the presented binning strategy. The adopted priors on the radial scale lengths are $R_\mathrm{disc}\sim\mathcal{N}(4\,\mathrm{kpc}, (3\,\mathrm{kpc})^2)$ and $R_{\sigma,r/z}\sim\mathcal{N}(10\,\mathrm{kpc}, (6\,\mathrm{kpc})^2)$, the prior for velocity dispersion of the outlier component is a normal distribution $\sigma_{{v}}\sim\mathcal{N}(100\,\mathrm{km\,s}^{-1}, (80\,\mathrm{km\,s}^{-1})^2)$ and the other priors are uniform as defined in the previous section.
The posterior distributions for the fits of each period bin are given in the supplementary material
and are summarised by the medians and percentiles in Table~\ref{MCMC_parameters}. The contamination fraction $\epsilon$ is generally small and $\sigma_{{v}}$ generally large for all period bins. Table~\ref{actual_params} reports the physical radial and vertical velocity dispersion normalization and scalelength in the solar neighbourhood, $\widetilde{\sigma}_{i,0}$ and $\widetilde{R}_{\sigma,i}$ respectively.

To verify the results of the MCMC fitting, we generate mock samples for the best-fit parameters according to the procedure from Section~\ref{sec:mock}, and we make use of the weights for the mock sample to compare the kinematics of the fitted model with the dataset under the same spatial distribution. In Fig.~\ref{v_l_v_b_distribution}, we have plotted the $v_\ell$ and $v_b$ distributions of these mock samples compared to that of the observations, where $v_{\ell/b} = s\cdot\mu_{\ell/b}$. We have chosen to omit the lowest period bin ($80-150$ days) from this plot and in later plots and analysis because the contamination rate, $\epsilon$ is the highest among other period bins (see Table~\ref{MCMC_parameters}) and it is likely it does not follow the broad trend of increasing dispersion with decreasing period due to contamination from short-period-red stars as we will discuss in Section~\ref{section::improvements}.
For the displayed period bins, the mock samples generally agree with the observations. For some period bins, the shape of the observed $v_{b}$ is sharper than the mock sample implying that our modelling has some caveats. Three reasons could lead to this: first, the assumed outlier distribution did not characterise the contamination accurately and underestimated the outlier star contribution consequently. Secondly, the period binning strategy needs to be improved. Bins at long periods cover Mira variables of a broader range of ages than the bins at short periods. Hence, if the younger stars in the period bin have much smaller velocity dispersion than the average of the bin, the sharper peak in observation would appear while the general shape of the overall distribution is still correct. Thirdly, the assumed functional form for the velocity dispersion parameters $\sigma_{i} = \sigma_{i,0}\exp{(R_{0}-R)/R_{\sigma,i}}$ may be inappropriate. We illustrate this final possibility by plotting the radial profile of the longitudinal and latitudinal velocity dispersions $\sigma_{\ell}$ and $\sigma_{b}$ in Fig.~\ref{sigmal_sigmab_radial_profile}. For one or two period bins, the large $R$ radial behaviour of $\sigma_{b}$ is not completely in agreement with the observations. The $\sigma_{\ell}$ distribution is relatively more poorly fitted than the $v_{b}$ distribution. Again, this could be due to the adopted form of the distribution function. However, apart from these very minor discrepancies, our modelling is in agreement with the observations. This is reinforced by the comparison of the $v_{\ell}$ and $v_b$ distribution for $275<\mathrm{Period/day}<300$ in Fig.~\ref{vl_vb_radial_profile}. The model is in good agreement with the observations. We will discuss further limitations of our approach in Section~\ref{sec::discussion}. 

As noted previously, the spatial distribution of stars has not been considered in the modelling as it is subject to completeness effects arising from Gaia's scanning law and the effects of extinction. As a result, the spatial distribution of the (unweighted) mock samples and the observations are in disagreement when the completeness of the dataset is not considered. Our weighting of the mock samples reproduces the spatial distribution of the data enabling comparison of the kinematic fits as shown in Fig.~\ref{v_l_v_b_distribution}, for example. When the weights are not considered, the mock sample distribution can be considered as the approximate underlying completeness-corrected distribution of the data (only up to a point as according to our procedure, where there is no data there will also be no mock stars). The weights are thus giving the proportion of stars at each $\bs{x}$ that have been observed. This is demonstrated in Fig.~\ref{vertical_density_profile} by comparing the unweighted Galactic height distribution of the mock sample to the dataset. Note that our procedure only gives access to the relative completeness so the histograms have been chosen to be normalized. The distributions of the data points are generally broader than the unweighted mock distributions, which we interpret as incompleteness in the dataset towards the Galactic midplane, possibly arising from extinction. 
This interpretation of the unweighted mock samples assumes the distribution functions well describe the Milky Way sub-populations. We discuss the shortcomings of the approach later, but the good agreement in Fig.~\ref{vertical_density_profile} also demonstrates that even without considering incompleteness, the distribution functions do a good job of describing the data.

\section{Period--age relationship}\label{sec::period_age_relation}
\begin{figure}
    \centering
    \includegraphics[width =\columnwidth]{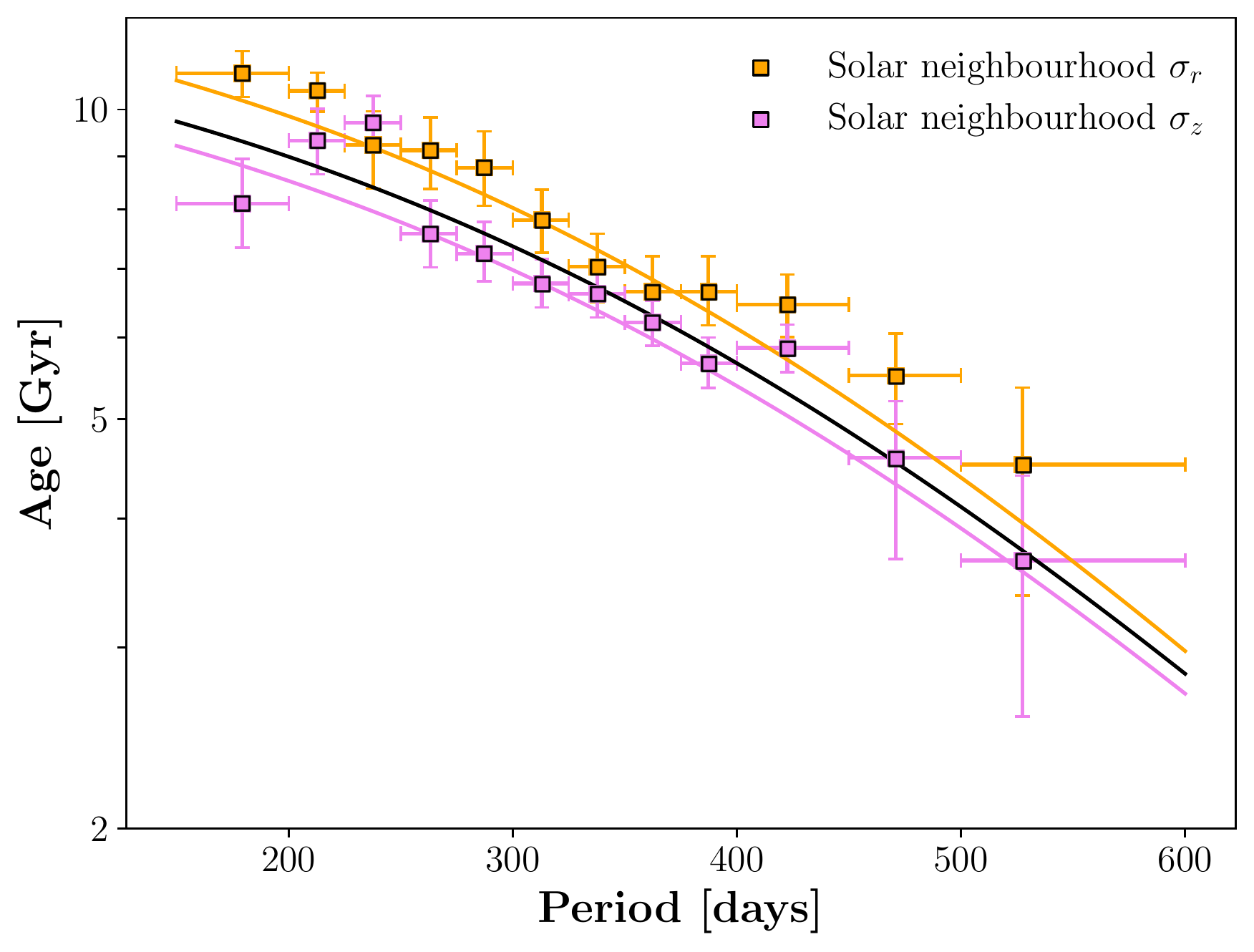}
    \caption{The calibrated age--period relationship of the O-rich Mira variables. The orange and violet points are the velocity dispersion from the kinematic modelling. The orange, purple, and black lines are the fitted period--age relations using radial, vertical velocity dispersions, and two together respectively, with fitted parameters given in Table~\ref{table:functionalform}.}
    \label{age-period}
\end{figure}

With the dynamical distribution functions in each Mira variable period bin well characterised, we now turn to what this implies for the corresponding age of each period bin. To do this we must adopt an age--velocity dispersion relation (AVR). We choose the AVR measured by \cite{Yu_2018} from LAMOST data of $\sim3500$ sub-giant/red giant stars. \cite{Yu_2018} characterised the velocity dispersions of their sample split into age bins using the entirety of their dataset and also for two sets split by Galactic height: $|z|<0.27\, \mathrm{kpc}$ and $|z|>0.27\,\mathrm{kpc}$. The ages of stars in \cite{Yu_2018} were estimated by comparing the stellar parameters ($[\mathrm{Fe/H}]$, $\mathrm{T_{eff}}$, $\log g$) measured by LAMOST to a grid of isochrone models. Age estimates were found by marginalizing the likelihood over initial mass and absolute magnitude. The AVRs were produced by further binning stars in their sample by age. This procedure accounts for uncertainties arising from the velocities but \emph{not} the ages. We discuss the impact of this later. 

We estimate the corresponding AVR of our sample by averaging the two $|z|$-separated AVRs in \cite{Yu_2018} weighted by the number of stars in our sample that are above and below $|z| = 0.27 \,\mathrm{kpc}$ in each bin. Consequently, the final AVR was slightly different for each bin. At low ages, the corresponding AVR is not monotonic due in part to uncertainties and the low numbers of stars in some low-age bins. Thus, we remove points in the AVR if the age is less than that of the previous age bin so that we could interpolate a monotonic AVR to find an age at each radial and vertical dispersion, $\widetilde{\sigma}_{r,0}$ and $\widetilde{\sigma}_{z,0}$. The uncertainty is again propagated using Monte Carlo samples. The final calibrated age--period relationship is shown in Fig.~\ref{age-period}. 

\begin{figure*}
    \centering
    \includegraphics[width = \textwidth]{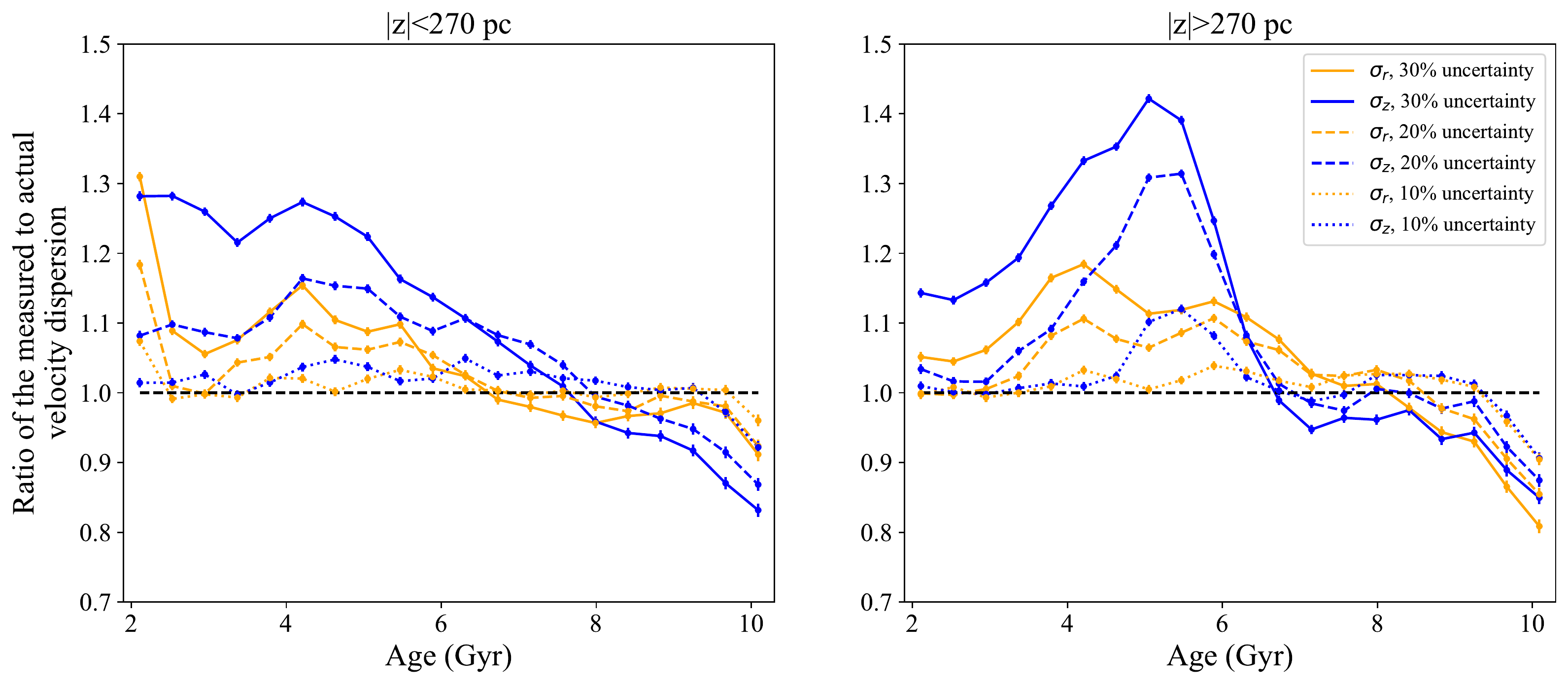}
    \caption{The ratio of the age--velocity dispersion relation broadened by different age uncertainties ($10, 20$ and $30\percent$ denoted by dotted, dashed and solid) relative to the `true' age--velocity dispersion relation without age uncertainties. The left panel shows results for the  $|z|<0.27 \,\mathrm{kpc}$ AVR from \protect\cite{Yu_2018} and the right panel their age--velocity dispersion relation for $|z|>0.27 \,\mathrm{kpc}$. Yellow lines correspond to $\sigma_r$ and blue $\sigma_z$.}
    \label{correction_factor}
\end{figure*}

\begin{figure*}
    \centering
    \includegraphics[width = \textwidth]{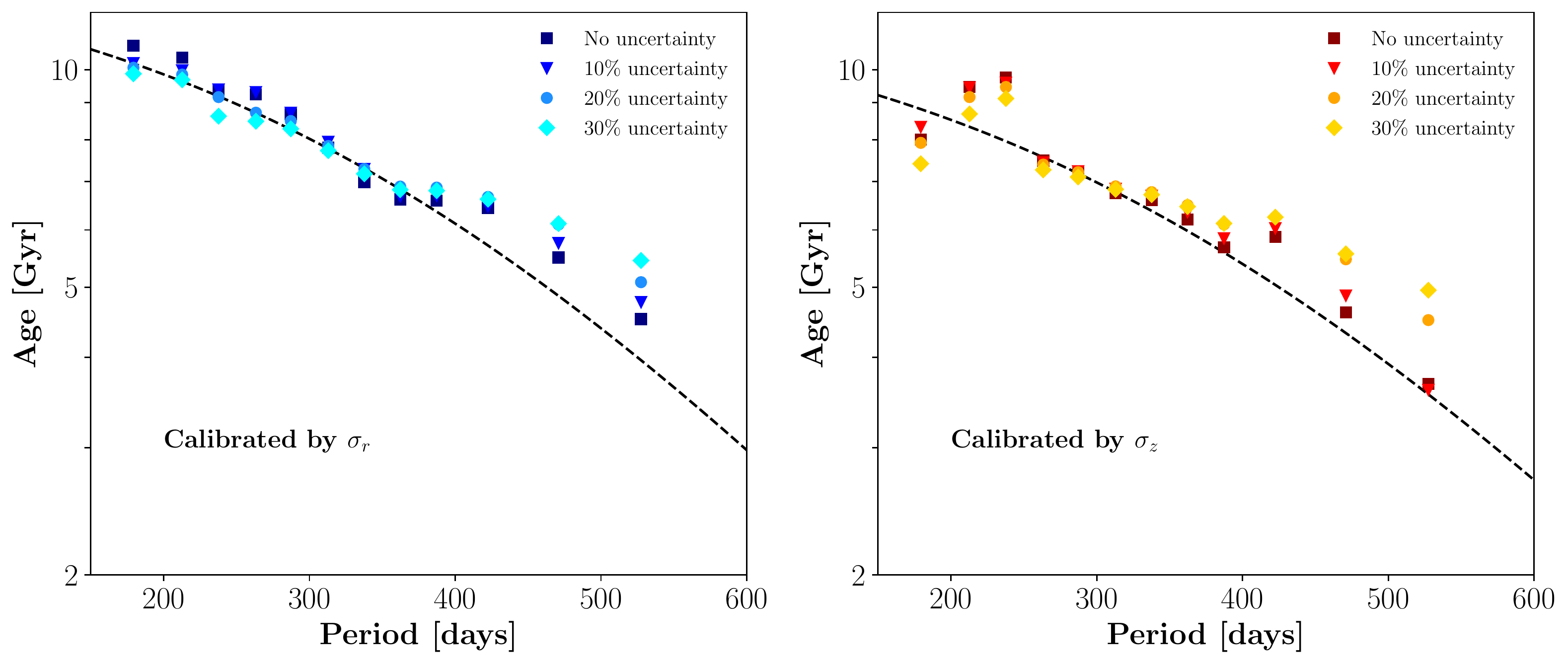}
    \caption{The calibrated period--age relationship using age--velocity dispersion relations broadened by different age uncertainties (as labelled in the legend). The relation calibrated by $\widetilde{\sigma}_{r,0}$ is shown in the left panel while $\widetilde{\sigma}_{z,0}$ is on the right. The error bars are not shown in this figure. The black dashed lines in both panels are the fitted period--age relations shown by the orange and pink lines in Fig.~\ref{age-period} respectively.}
    \label{comparing_age_period_relation}
\end{figure*}

\cite{Yu_2018} discussed that the uncertainties in the estimated ages of stars would broaden the measured AVR. \cite{Liu_2015} argued that the age estimation method used in \cite{Yu_2018} could have uncertainties at the $30\percent$ level which propagate from the uncertainties of the LAMOST stellar parameters. Here, we will discuss how much this effect would affect the period--age relationship. We generate $500\,000$ stars with uniformly distributed ages and assign each star a radial and vertical velocity from a Gaussian distribution centred at $0$ with standard deviations of $\sigma_{r}$ and $\sigma_{z}$ calculated from the AVR. Then, the ages of the stars are scattered by $(10, 20, 30)\%$ uncertainties. We then bin the stars with the scattered age and calculate the measured radial and vertical velocity dispersion. The ratio of the measured to actual velocity dispersion for the AVR is given in Fig.~\ref{correction_factor}, where the left and right panels are made for the AVR of $|z|<0.27 \,\mathrm{kpc}$ and $|z|>0.27 \,\mathrm{kpc}$ respectively. We divide this ratio by the corresponding velocity dispersions in the AVR as a correction. In Fig.~\ref{comparing_age_period_relation} we show the period--age relations calibrated using AVRs with different levels of age uncertainty. We see that with $30\%$ uncertainty in AVR the maximum correction could be up to $20\%$ in age as calibrated from $\sigma_{R,0}$ and $34\%$ from $\sigma_{z,0}$.

We have also considered other recent AVR calibrations available in the literature. For example, \cite{Sharma2021} have provided a fit of the radial and vertical dispersions in a separable form in terms of the age, angular momentum, metallicity and Galactic height. Their relations produce significantly smaller dispersions at fixed age such that the derived period--age relation will assign significantly larger ages at fixed period which in the extreme can be $\gg14\,\mathrm{Gyr}$. We are therefore inclined to use the \cite{Yu_2018} relations and the applicability of the \cite{Sharma2021} relations merits further investigation.

\section{Discussion}\label{sec::discussion}
We now turn to the interpretation and understanding of our results, in particular concentrating on the comparison with previous period--age estimates for Mira variable stars and possible future model improvements.

\subsection{Comparison with Mira variable cluster members and previous results}
\begin{figure*}
    \centering
    \includegraphics[width=\textwidth]{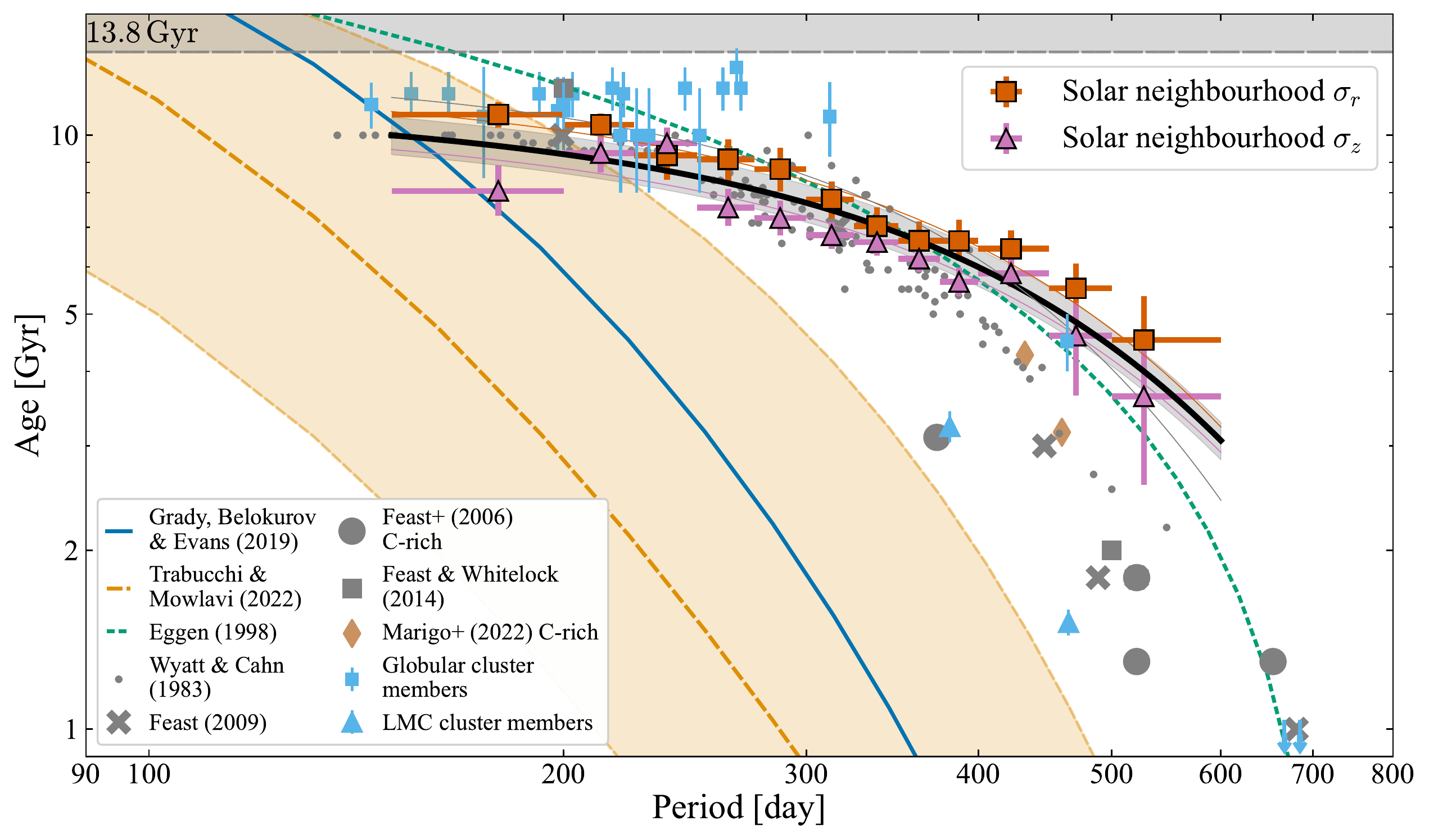}
    \caption{Comparison of the derived period--age relations with other literature results. The orange squares and pink triangles show our Mira variable period--age measurements from Table~\ref{actual_params}. The small grey points are from the models of \protect\cite{WyattCahn1983}, the green short-dashed line from the model of \protect\cite{Eggen1998} and the orange long-dashed line from the model of \protect\citet[][along with the associated scatter shown by the shaded region]{Trabucchi2022}. The light blue squares are Mira variable globular cluster members from \protect\cite{Clement2001}, the brown diamonds C-rich Mira variable open cluster members from \protect\cite{Marigo2022} and the light-blue triangles LMC cluster members. The solid blue line is a fit from \protect\cite{Grady2019} to a broader sample of LMC cluster members. The grey points are period--age estimates for disc populations from \protect\cite{Feast2006}, \protect\cite{Feast2009} and \protect\cite{FeastWhitelock2014}. The black line is the joint fit of our results and the globular cluster members from Table~\ref{table:functionalform} and the thinner orange, pink and grey lines show the other three fits from that same table.}
    \label{fig:period-age-relationship}
\end{figure*}
In Fig.~\ref{age-period} we display a series of period--age indicators of Mira variable stars. The age--kinematic method for period--age calibration has been utilised by \cite{Feast2006}, \cite{Feast2009} and \cite{FeastWhitelock2014}. \cite{FeastWhitelock2000} demonstrated that Mira variables in the solar neighbourhood exhibited clear correlations between period and kinematics. These have been translated approximately into period--age measurements using results from the solar neighbourhood in the cited works. However, it should be said that all of the quoted results are only approximate due to the absence of robust age--kinematics calibrations.

Mira variables in clusters give a more direct measurement of the period--age relation than the indirect method using the age--kinematic calibrations. Unfortunately, there are comparatively few cluster Mira variables. Those in globular clusters have been studied by \cite{Sloan2010} whilst those with good evidence of Milky Way open cluster membership from Gaia have been studied by \cite{Marigo2022}. There are also many candidates for LMC cluster membership as studied by \cite{Grady2019}. However, membership of an LMC cluster is difficult to discern purely from projected coordinates \citep[as used by][]{Grady2019} and proper motion data. We compile Mira variable globular cluster members using the globular cluster variable star compilation from \cite{Clement2001}. We consider all stars flagged as `M' or `M?', and not flagged as a likely field star (`f' or `f?'). Furthermore, if available, we ensure the Gaia DR3 proper motion is within $3\sigma$ of the measured cluster mean proper motion from \cite{VasilievBaumgardt}. Here $\sigma$ is a quadrature sum of the measurement uncertainty and the central velocity dispersion. We complement with ages primarily from \cite{Vandenberg2013} and \cite{Dotter}, and from \cite{Beaulieu2003} for NGC 6553, \cite{Geisler} for Terzan 7, \cite{Ortolani} for Terzan 1, \cite{MarinFranch2009} and \cite{ForbesBridges} for NGC 6441 and \cite{SantosPiatti} for NGC 6356, NGC 6388, NGC 6642 and NGC 6760. Terzan 5 has evidence of multiple star formation events \citep{Ferraro2016} so we assign stars with periods $< 400$ day an age of $12\,\mathrm{Gyr}$ and longer-period stars an age of $4.5\,\mathrm{Gyr}$. There is a carbon-rich Mira variable in the old globular cluster Lyng\aa\ 7 that has been suggested as a product of binary evolution \citep{FeastLynga7}. However, its Gaia DR3 proper motion is not consistent with being a cluster member. Its radial velocity is perfectly consistent so one possibility is that the Gaia measurement is spurious. This seems quite likely as there are two nearby Gaia DR3 sources with only two-parameter astrometric solutions suggesting contamination in the Lyng\aa\ 7 C-rich Mira variable measurement.

For Mira variable open cluster members, we use the compilation from \cite{Marigo2022} adopting their measured periods and the cluster ages from \cite{CantatGaudin2020}. \cite{Marigo2022} identify some cluster members on the fundamental period--luminosity relation followed by Mira variable stars but with too low an amplitude for traditional Mira variable classification. We consider all stars that \cite{Marigo2022} identify as fundamental pulsators and with $G$ band amplitudes greater than $0.865\,\mathrm{mag}$ \citep{Grady2019} estimated from the photometric uncertainties. There are two such stars with are both C-rich.

Finally, we consider possible LMC and SMC cluster members from the Gaia DR3 LPV candidate catalogue. We combine the list of cluster ages from \cite{Baumgardt2013} and \cite{Bonatto2010}. To limit contaminants, we conservatively find all Gaia DR3 LPV candidates within one cluster radius as determined by \cite{Bica2008} (adopting the median cluster radius of $0.45\,\mathrm{arcmin}$ when a radius is not available). We further limit to those with proper motions within $3\sigma$ of $(\mu_{\alpha}*,\mu_\delta)=(1.910,0.229)\mathrm{mas\,yr}^{-1}$ \citep{LMCpropermotion} where $\sigma$ is the quadrature sum of the uncertainties and $100\,\mathrm{km\,s}^{-1}$ at the distance of the LMC, and those with distances between $30$ and $70\,\mathrm{kpc}$ as determined from equation~\eqref{plr}. We isolate Mira variables by restricting to stars with $G$ amplitudes $>0.865\,\mathrm{mag}$ as determined by the $G$ photometric uncertainties and the Fourier light curve fits. This results in $4$ high-confidence LMC cluster members.

The described combination of cluster measurements is shown in Fig.~\ref{fig:period-age-relationship}. We see in general the good agreement between the results derived from the age--kinematic relation and the cluster members. There are some globular cluster members with longer periods but higher ages (most notably the $312$ day period Mira in NGC 5927 which has an age of $12.25\,\mathrm{Gyr}$ from \citealt{Dotter} and $10.75\,\mathrm{Gyr}$ from \citealt{Vandenberg2013}). This may reflect metallicity dependence in the period--age relation or these could be the results of binary evolution in these clusters producing slightly more massive AGB stars than expected at fixed age.

There are several theoretical period--age relations from the literature. The earliest of these are the results from \cite{WyattCahn1983} who found ages for local Mira variable stars via main-sequence mass estimates derived from models of Mira variables as fundamental pulsators which were fitted to optical and infrared photometry and periods. \cite{Eggen1998} similarly provided a theoretically-motivated period--age relation by supposing fundamental Mira-like pulsations occur once a star of a given mass (age) reaches some critical radius. Most recently, \cite{Trabucchi2022} have used theoretical models to produce period--age calibrations for O-rich and C-rich Mira variable stars. They highlighted one expectation of the models is a large spread of age at fixed period. Furthermore, their period--age relations agreed very well with the cluster member measurements mostly compiled by \cite{Grady2019}. However, as we have hinted at above, there is perhaps good reason to believe that the LMC cluster members are quite a contaminated set and that LMC field stars coincident on the sky with the clusters are likely to be incorrectly identified as cluster members. The field stars will typically be older than the cluster members, having already left their parent clusters, and so these contaminants will act to decrease the typical age at fixed period. It could be that there is an additional variable controlling the period--age relation that produces the discrepancy between the LMC clusters and the local age--kinematic relations. The spread in models from \cite{Trabucchi2022} is almost consistent with the measurements made here. However, the globular clusters suggest any metallicity dependence would go the other way. Furthermore, binary evolution produces higher periods at fixed age so would not explain the discrepancy.

A further supporting piece of evidence for the age--period relation we have derived here is the properties of the LMC population as a whole and the Galactic bulge sample. In both sets, there are stars with $\sim500-600\,\mathrm{day}$ periods. From our calibrations, these stars are $\sim3-4\,\mathrm{Gyr}$ old. The LMC has a tail towards longer-periods consistent with even more recent star formation. The Galactic bulge is primarily considered as an old population \citep{Zoccali2003} although there has been significant evidence that there are intermediate-age populations as young as $3\,\mathrm{Gyr}$ \citep{Bensby2013,Bensby2017,Nataf2016}. Our calibration is entirely consistent with these results. A lower age--period relation would mean a significant population of stars in the Galactic bulge with $\lesssim1\,\mathrm{Gyr}$ old populations although again we should stress the expected spread in ages at each period could still produce some consistency in the results.

\subsection{A parametric period--age relation}
Our fitting has provided the approximate ages of O-rich Mira variable populations in a series of period bins. It is more convenient to work with an analytic relation that approximately fits the results. The flexible form
\begin{equation}
\tau = \tau_0\frac{1}{2}\left(1+ \tanh\Big[\frac{330-P(\mathrm{days})}{P_s}\Big] \right),
\label{eqn::period-age-relationship}
\end{equation}
provides an approximate fit to the data. We take the data reported in Table~\ref{actual_params} and fit equation~\eqref{eqn::period-age-relationship} allowing for an additional fractional scatter in the ages of $f_\tau$ such that the age errors are $\sqrt{\sigma_\tau^2+f_\tau^2\tau^2}$. $(\tau_0,P_s,f_\tau)$ are given logarithmic flat priors and we sample using \textsc{emcee} \citep{emcee}. We fit for $\sigma_r$ and $\sigma_z$ both separately and jointly and report the results in Table~\ref{table:functionalform}. Although the dispersion parameters are derived from the same model fit, the corner plots in the supplementary material 
demonstrate the parameter constraints are uncorrelated for nearly all period bins validating treating the results in this way. We also perform a joint fit of the dispersion results together with the globular cluster member compilation described in the previous section, again reporting the results in Table~\ref{table:functionalform}. All four sets of results are quite consistent with $\sigma_z$-only fits producing the lowest age at fixed period and the combination with the globular clusters producing the highest. As expected, the scatter is largest for the combined fit with the globular clusters but nevertheless, the scatter is only around $10\percent$ in age.

\begin{table}
    \caption{Functional form for the period--age relation fitted to our results. We adopt the form $\tau = (\tau_0/2)\left(1+ \tanh((330-P(\mathrm{day}))/P_s\right)$ with a fractional age uncertainty of $f_\tau$.}
    \centering
    \begin{tabular}{cccc}
Subset&$\tau_0$&$P_s$&$\ln f_\tau$\\
\hline
$\sigma_r$&$14.9\pm0.7$&$389\pm77$&$-6.70\pm0.01$\\
$\sigma_z$&$13.0\pm0.5$&$404\pm111$&$-5.71\pm0.03$\\
Both&$13.7\pm0.6$&$401\pm88$&$-2.63\pm0.04$\\
With GC&$14.7\pm0.7$&$308\pm54$&$-2.17\pm0.04$\\
    \end{tabular}
    \label{table:functionalform}
\end{table}

\subsection{Model limitations and future improvements}\label{section::improvements}
Before concluding, we will discuss some of the limitations of our modelling and possible improvements that could be adopted in future analyses.

\emph{Binning in period}: We have opted to bin our data in period and analyse each period bin independently. This is a valid approach as the period uncertainties are typically quite small: the median period uncertainty is $11.6$ day and $7.1$ day for $\mathrm{period}<300\,\mathrm{day}$. Hence, our strategy is valid for most of the period bins considered. A further generalization is to express the models in terms of the period as a continuous subpopulation label. We then have to introduce hyper-parametrizations for the parameters in $f(\bs{J})$ to express $f(\bs{J}|P)$. The integrals would involve an additional integral over the label $P$ and we would have a weighting of the populations $f(P)$ (which if we are considering periods as proxies for age is akin to a star formation rate and could be an exponential in age, for example). The advantage of this approach is a more principled accounting of the period uncertainties as well as providing a route to consider the spread in age (kinematics) at each period that might arise from helium flashes, hot-bottom burning or the presence of short-period red stars. The downside of such an approach is that we would have to fit a parametrized form for the parameters as a function of $P$ making the models significantly more complicated and potentially producing biased by our choice of functional form.

\emph{Velocity dispersion profile}: We have here adopted a simple pure exponential decay for the velocity dispersion of each period bin. This form gives a good fit of the models to the data, particularly as we have chosen a rather limited Galactocentric radial range. It has been suggested that the velocity dispersion in the outer disc flattens or even increases with radius \citep{SandersDas2018,Mackereth2019}. For example, \cite{Sharma2021} argues that the pure exponential decay of the velocity dispersions is not well motivated by the data, which shows signs of a rising dispersion beyond the solar radius. To incorporate this possibility, one possible change is to modify $\tilde\sigma_i(R_c)$ as
\begin{equation}
    \tilde\sigma_i(R_c) \equiv \sigma_{i,0} (\exp[ -(R_c-R_0) / R_{\sigma,i} ] +\alpha_i(R_i/R_0)^2)/(1+\alpha_i),
    \label{alpha_term}
\end{equation}
with the additional fitting parameters $\alpha_i$ to match the flattening/upturning dispersion profiles in the outer disc as suggested by \cite{Sharma2021}. This may be a necessary enhancement when modelling the data beyond the extended solar neighbourhood. For example, if one were to consider investigating possible metallicity dependence of the period--age relation. However, such an enhancement does not seem necessary for our data.

\emph{Limitations of equilibrium axisymmetric distribution function approach}: It is reassuring to note that the age estimates from the radial and vertical dispersions separately give very similar results for the period--age relation of the O-rich Mira variables. However, the relation derived from the radial dispersion is consistently higher than that derived from the vertical dispersion. We have seen how our dynamical models capture well both the longitudinal and latitudinal velocity distributions of the sample but typically the latitudinal distributions are better modelled suggesting our results are more reliable for the period--age relation derived from $\sigma_z$. This occasional mismatch of the longitudinal dispersion in Fig.~\ref{sigmal_sigmab_radial_profile} could be a shortcoming of the use of a quasi-isothermal distribution function. There are other action-based disc models available in the literature \citep[e.g.][]{Binney2022} which could be explored. As mentioned previously, using a dynamical distribution function simply incorporates the required asymmetry in the azimuthal component as well as necessarily linking together the radial and azimuthal dispersions due to the requirement of dynamical equilibrium. There could also be inconsistencies arising from this assumption of equilibrium as it is known that the Galactic disc shows non-equilibrium structure at the $5-10\percent$ level. Any inflation of the velocity dispersion as a result of this is not a concern as we have anchored to tracers that will also display this inflation. The assumption of axisymmetry could also be giving rise to similar variations. We are using the velocity dispersion at the solar radius from a range of different azimuths but if the velocity dispersion is varying significantly with azimuth \citep[e.g.][]{DrimmelGaia}, the comparison between our sample and the age--velocity dispersion results from \cite{Yu_2018} may be inappropriate. Furthermore, our model has assumed a fixed Milky Way potential from \cite{McMillan2017}. Whilst this potential captures many of the global features of the Milky Way, it may not in detail be appropriate across the entirety of the Galactic disc region considered here. In the wrong potential, it may be very difficult to fully match the full velocity distribution of the data at every spatial location. Reasonable variations of the potential will likely inflate the uncertainties in our derived parameters. We should also note that although we have inflated the Gaia astrometric uncertainties in our analysis to reflect shortcomings of the current Gaia data processing, it is likely that future Gaia data releases will improve the uncertainty estimates providing a better handle on the underlying dispersions of the disc populations. This may decrease the dispersion for the youngest populations (e.g. the $500-600$ day period bin) but the dispersions of the oldest populations are very insensitive to the uncertainties so we believe our measurements are reliable.

\emph{SP-red stars}: We found the stars in our lowest considered period bin ($80-150$ day) have significantly lower dispersions and hence lower ages than the neighbouring $150-200$ day bin (see Table~\ref{MCMC_parameters}). This bucks the broad trend seen in e.g. Fig.~\ref{fig:period-age-relationship} and for this reason, as well as the fact that this bin requires the largest outlier fraction of all modelled bins, we decided to neglect these results in our period--age relation fits. \cite{FeastWhitelock2000} found a similar effect from Hipparcos data that they attribute to short-period(SP)-red stars which contaminate the short-period end and are kinematically more similar to the longer-period Mira variables. It is not clear exactly what the origin of these stars is and they could represent a different evolutionary stage to the bulk Mira variable population. \cite{FeastWhitelock2000} hypothesise they could be stars on their way to becoming longer-period Mira variables or temporarily dimmed during their helium-shell flash cycle \citep{Trabucchi2017}. From Gaia-2MASS colour-colour diagrams, we did not clearly identify a distinct population of SP-red-like stars in the short-period bin but it is likely they are present and potentially also more weakly contaminating the $150-200$ day bin which also shows a slightly lower $\sigma_z$ than perhaps expected. It is known that Mira variables in globular clusters follow a period--metallicity relation with shorter-period stars more prevalent in metal-poor clusters \citep{FeastWhitelock2000P}. This then suggests that the shortest period bin we considered has significant contamination from more metal-poor objects and is not representative of the broader solar neighbourhood samples used to calibrate the period--age relations. However, it is then surprising that a more metal-poor population would have a lower than expected dispersion as in both in-situ and accreted scenarios the opposite is likely the case. More generally, our methodology could be impacted by metallicity effects. We have already limited to O-rich Mira variables which should preferentially remove metal-poor stars. Further investigation is required to separate out the degeneracies between period, age and metallicity, and a possible avenue is to consider the variation of kinematics with unextincted colour as a metallicity proxy \citep[e.g.][]{Alvarez1997}.

\emph{Hot-bottom burning}: 
From equation~\eqref{plr}, the slope of the period--luminosity relation changes after $\mathrm{Period (days)}>400$. This hints that our O-rich Mira variable star sample with periods above $400$ days is a mixture of hot-bottom burning (HBB) stars and low-mass fundamental pulsators right at the end of their lifetime \citep{Whitelock2003,Trabucchi2019}. The balance of these two kinematically distinct populations depends on the star-formation history (e.g. the HBB population would be reduced if there is no recent star formation). Hence, as we are measuring the average age at a fixed period, our result is somewhat related to the star-formation history of the Milky Way. This mixing of HBB stars likely also broadens the period--age relation for $\mathrm{Period (days)}>400$ \citep[as it perhaps does the period--luminosity relation e.g.][]{Ita2011}, and it might address the small discrepancy between our relation and the literature results shown in Fig.~\ref{fig:period-age-relationship}. We hypothesise that the period--age relation is more universal and reliable for periods under $400$ days.

\emph{C-rich stars}: Finally, a further direction is to consider the C-rich Mira variables from Gaia. C-rich Mira variables also follow period--luminosity relations that are typically broader than that for the O-rich Mira variables due to circumstellar dust \citep{Ita2011}. They also appear to trace period--age relations \citep[e.g.][and evidenced in Fig.~\ref{fig:period-age-relationship}]{Feast2006}. Typically they are less abundant in the Galaxy than the O-rich counterparts \citep{Ishihara2011} but importantly are biased towards younger ages \citep[and lower metallicities, e.g.][]{Boyer2013} so present a route to better constraining the longer-period end of the Mira variable period--age relation.

\section{Conclusions}\label{sec::conclusions}
We have used the Gaia DR3 long-period variable candidate catalogue to produce a calibration of the Mira variable period--age relation. Using a carefully selected population of likely O-rich Mira variable stars, we have fitted a series of action-based dynamical models to the stars separated by period. We have found very good model fits for the velocity distributions of our sample from which we have derived period--kinematic relations for the solar neighbourhood. Comparison with an age--velocity dispersion relation for sub-giant/red giant stars in the solar neighbourhood has allowed us to provide a calibration of the Mira variable period--age relation.

Our derived relation agrees well with previous literature approaches using a similar methodology and with the members of clusters with known ages. Some theoretical models agree well with the derived relation but more recent calibrations appear to be consistently younger at fixed period than our relations suggest. Consideration of the age distribution of Mira variable stars in the Galactic bar-bulge produces a consistent picture with other bar-bulge age tracers using our relation.

This new period--age relation opens the possibility of inspecting the star formation history and evolutionary properties of distant and/or highly-extincted regions of our Galaxy and the Local Group. Mira variables are some of the brightest stars in an intermediate-age population, their infrared brightness makes them ideal tracers of dusty environments, and their high amplitude and long periods mean they suffer low contamination. For these reasons, in the era of JWST, Mira variables will provide us with a new window of the evolution of the Universe.

\section*{Data availability}
All data utilised in this work are in the public domain. In the supplementary material, we provide corner plots showing the posterior distributions of the dynamical model parameters for each period bin.

\section*{Acknowledgements}
We thank the anonymous referee for a careful reading of the paper and thoughtful comments that improved the presentation.
JLS thanks the support of the Royal Society (URF\textbackslash R1\textbackslash191555).
This work has made use of data from the European Space Agency (ESA) mission
{\it Gaia} (\url{https://www.cosmos.esa.int/gaia}), processed by the {\it Gaia}
Data Processing and Analysis Consortium (DPAC,
\url{https://www.cosmos.esa.int/web/gaia/dpac/consortium}). Funding for the DPAC
has been provided by national institutions, in particular, the institutions
participating in the {\it Gaia} Multilateral Agreement. This publication makes use of data products from the Two Micron All Sky Survey, which is a joint project of the University of Massachusetts and the Infrared Processing and Analysis Center/California Institute of Technology, funded by the National Aeronautics and Space Administration and the National Science Foundation. This paper made use of
\textsc{numpy} \citep{numpy},
\textsc{scipy} \citep{scipy},
\textsc{matplotlib} \citep{matplotlib}, 
\textsc{seaborn} \citep{seaborn},
\textsc{pandas} \citep{pandas1},
\textsc{corner} \citep{corner}
\textsc{astropy} \citep{astropy:2013,astropy:2018},
\textsc{galpy} \citep{Bovy2015},
and \textsc{Agama} \citep{Agama}.

\bibliographystyle{mnras}
\bibliography{bibliography}

% Don't change these lines
\bsp    % typesetting comment
\label{lastpage}
\end{document}